\documentclass{article}
\usepackage{graphicx} 
\usepackage{amsmath,amssymb,amsthm}
\usepackage[utf8]{inputenc}
\usepackage[hidelinks]{hyperref}
\usepackage[capitalise,noabbrev]{cleveref}

\usepackage{pythonhighlight}

\usepackage{tikz}
\usetikzlibrary{quantikz2}

\usepackage[maxbibnames=15, url=false, style=alphabetic, bibencoding=utf8, giveninits=true]{biblatex}
\DeclareSortingNamekeyTemplate{
  \keypart{
    \namepart{family}
  }
  \keypart{
    \namepart{prefix}
  }
  \keypart{
    \namepart{given}
  }
  \keypart{
    \namepart{suffix}
  }
}
\title{Theoretical and experimental analysis of adaptive quantum computers}
\author{Niels M. P. Neumann$^{1,2}$}
\date{$^1$ Netherlands Organisation for Applied Scientific Research (TNO), The Hague, The Netherlands \\
$^2$ National Research Institute for Mathematics and Computer Science (CWI), Amsterdam, The Netherlands}

\newcommand{\R}{\mathbb{R}}
\newcommand{\F}{\mathbb{F}}
\newcommand{\eps}{\varepsilon}

\newcommand{\bigo}{\mathcal{O}} 
\newcommand{\LAQCC}{\mathsf{LAQCC}}

\newcommand{\floor}[1]{\left\lfloor #1 \right\rfloor}
\newcommand{\ceil}[1]{\left\lceil #1 \right\rceil}

\newtheorem{theorem}{Theorem}

\begin{document}

\maketitle

\begin{abstract}
    Fault-tolerant quantum computations require alternating quantum and classical computations, where the classical computations prove vital in detecting and correcting errors in the quantum computation. 
    Recently, interest in using these classical computations has been growing again, not to correct errors, but to perform computations. 
    Various works have looked into these so-called adaptive quantum algorithms.
    Few works however have looked in the advantages of adaptive quantum algorithms in realistic scenarios. 
    This work provides the first step in this direction. 
    We introduce a worst-case noise model and use it to derive success probabilities for preparing a GHZ state and preparing a $W$-state using either an adaptive quantum algorithm, or using a standard non-adaptive quantum algorithm. 
    Next, we implemented these protocols on quantum hardware and we compare the outcomes to our derived theoretical results. 
    We find that despite their potential, adaptive quantum algorithms currently do not outperform full quantum algorithms. 
\end{abstract}

\section{Introduction}
Recent results have shown improved error rates by applying error correcting codes~\cite{GoogleWillow:2024,Hong:2024,Daguerre:2025,Dasu:2025}.
Most error correction routines work by computing some syndrome of a group of qubits, measuring the syndrome, computing correction terms on a classical computer, and applying these correction terms. 
Error correction thus uses interactions between a quantum computer and a classical computer. 
Recently, this interaction has received newfound interest, not for correcting errors, but to enhance computational power instead. 

\citeauthor{PhamSvore:2013} showed how intermediate classical computations can help factor integers in polylogarithmic depth~\cite{PhamSvore:2013}. 
The main tool in their work was an efficient implementation of the quantum fanout gate, previously introduced by \citeauthor{HoyerSpalek:2005}~\cite{HoyerSpalek:2005}.
\citeauthor{Browne:2011} showed how this quantum fanout gate can be implemented using intermediate measurements and log-depth classical computations~\cite{Browne:2011}. 
Their construction effectively opened the way to implementing the quantum fanout gate on any quantum backend that supports intermediate classical computation. 

Later, \citeauthor{PiroliStyliarisCirac:2021} looked into implementing unitary operations using local quantum circuits and classical computations~\cite{PiroliStyliarisCirac:2021}. 
In this so-called $\mathsf{LOCC}$ model, they measure auxiliary qubits and use the measurement results to control future local unitaries. 
Their focus was on matrix-product states and which states are equivalent in this model. 
\citeauthor{Tantivasadakarn:2024} used similar ideas to topological states~\cite{Tantivasadakarn:2023,Tantivasadakarn:2024}.

In the remainder of this work, we call quantum algorithms that use intermediate calculations to control future quantum operations \textit{adaptive quantum algorithms}, and the computers implementing them \textit{adaptive quantum computers}.
The adaptive algorithms encountered so far only use feedforward of the measurement results or simple parity sum computations on the measurement results. 
Additionally, the depth of the quantum parts of the algorithms typically depends on the number of measurement rounds. 

\citeauthor{Buhrman:2024} introduced the \textit{Local Alternating Quantum-Classical Computations} ($\LAQCC$) model that explicitly bounds the number of communication rounds and the intermediate classical computations, thereby generalizing previous works~\cite{Buhrman:2024,Neumann:2025}. 
They showed for instance how various often-used quantum states, such as the GHZ state, the $W$ state, and the Dicke state, can be prepared in constant quantum depth with only a constant number of intermediate log-depth classical computations. 
Their work also used more complex intermediate classical computations, such as ordering. 

Adaptive quantum computers offload part of their computations to a classical computer, reducing the quantum resource requirements. 
The current age of noisy quantum computers necessitates smart use of the available quantum resources, as otherwise the output of the computation becomes completely uncorrelated from the intended one.
Arguments based on the required quantum resources do hint towards an advantage of adaptive quantum computers over non-adaptive quantum computers. 
Still, depending on the exact algorithm, such arguments only give an indication of the exact advantage offered by adaptive quantum circuits. 

As an example, when preparing a GHZ state on $n$ qubits, a standard approach using an all-to-all connectivity uses $n$ qubits, $n$ quantum gates, and has depth $\ceil{\log(n)}+1$, for a total circuit size of $\bigo(n\log n)$.
The adaptive approach by \citeauthor{Buhrman:2024} instead uses $2n-1$ qubits, around $4n$ quantum gates, and has constant depth, for a total circuit size of $\bigo(n)$. 
Based on the number of qubits required and the number of quantum gates applied, the standard approach seems favorable. 
On the other hand, based on the circuit size, an adaptive circuit seems favorable. 
The adaptive circuit is significantly more dense, meaning that qubits are idle only briefly. 
The approach that works best thus depends on the exact noise levels of the different terms in the algorithm. 

This work provides a thorough comparison of adaptive versus non-adaptive state preparation protocols. 
For two often-used quantum states we compare an adaptive quantum algorithm with a non-adaptive one and determine when one would outperform the other. 
These results typically depend on the underlying hardware and thus on the error model corresponding to the hardware. 
Common noise models, such as single- and two-qubit gate errors, read-out errors, and the dephasing and depolarizing channels, each focus on specific imperfections in quantum hardware. 
Typically, the more complex the noise model, the better it approximates the actual behavior of a quantum system. 

This work uses a simple and strict error model, where any error is assumed to produce the incorrect quantum state. 
Consequently, we have to assume a continuous gate set.
Discrete but universal gate sets approximate the quantum state to within arbitrary precision $\eps>0$. 
However, in the used error model, we cannot distinguish between approximation imprecision and errors. 
Even though this error model most likely overestimates the impact of errors, it does help in obtaining a first-order estimate of when adaptive quantum algorithms outperform non-adaptive quantum algorithms. 

To test our derived success probabilities, we test them on quantum hardware. 
Implementing quantum algorithms on hardware, especially adaptive ones, puts strain on the available quantum resources. 
Manually optimizing the implementation can help improve overall fidelity, but requires significant efforts. 
For our implementation, we therefore used the IBM Brisbane superconducting quantum backend~\cite{IBMQuantumExperience:2024}.
Even though the intermediate classical computations are limited to directly controlling future quantum operations by measurement results, this implementation does give guidance on the potential of adaptive circuits. 

\textbf{Summary of results}
\begin{itemize}
    \item This work provides a theoretical analysis of adaptive and non-adaptive protocols to prepare the GHZ state and the $W$-state. 

    \item Based on a first-order estimate using a worst-case error model, we find that an adaptive protocol exponentially outperforms a non-adaptive protocol if the probability of a CNOT-gate introducing an error is at most the probability that an error is introduced while a qubit is idling for $\Omega(\log n)$ CNOT-gates, resp., $\Omega(n)$ CNOT-gates, when using an all-to-all or linear nearest-neighbor connectivity, see also \cref{thm:error:GHZ}. 

    \item Implementations of the approaches on superconducting quantum hardware show that in practice protocols perform similar for larger $n$, with the effect of noise dominating the results. For larger $n$, we find a low success probability, supporting the degrading results, see also \cref{sec:hardware_implementation}. 

    \item A similar analysis of the $W$ state protocol shows that adaptive circuits perform better than non-adaptive circuits if the probability of a CNOT-gate introducing an error is at most probability that an error is introduced while a qubit is idling for $\Omega(n/(\log (n)\log\log (n)))$ CNOT-gates, see also \cref{thm:error:W_state}. 
\end{itemize}

\textbf{Organization of the paper}
This work presents an in-depth theoretical analysis of adaptive quantum computers and conducts experiments to test these theoretical derivations. 
We derive success probabilities for adaptive quantum state preparation protocols to prepare the GHZ state and the $W$ state from~\cite{Buhrman:2024} and compare their performance with standard non-adaptive protocols. 
First, \cref{sec:error_model} introduces the used error model. 
Then, \cref{sec:GHZ} provides the theoretical derivation for the GHZ state protocols. 
We consider a standard approach using an all-to-all connectivity and using a linear nearest-neighbor connectivity, and we consider an adaptive approach.
Additionally, we also derive expressions for the success probability of a hybrid version of a standard approach and an adaptive approach.
We then implement the protocols on quantum hardware to estimate their performance in \cref{sec:hardware_implementation}. 
We repeat this analysis in \cref{sec:W_state} for the $W$ state, and we conclude in \cref{sec:discussion} with a discussion and directions for future research.

\section{Error model}\label{sec:error_model}
Error models describe the behavior of quantum systems in noisy environments. 
Different error models exist and they typically assume some underlying physical behavior. 
Common noise models include single- and two-qubit gate errors, read-out errors, depolarizing noise, and dephasing noise. 
Naturally, a trade-off exists between the complexity of the noise model and the correspondence with the actual behavior. 

For our derivations, we use a worst-case error model where every error corresponds to a Haar-random unitary applied to the qubit(s) instead of the intended quantum gate.
For idling qubits, the intended gate corresponds to an identity operation. 

In this error model, the probability that two errors cancel each other is $0$. 
Suppose an error $B$ occurred and suppose a gate $G$ is applied after which another error $D$ occurs. 
The probability that the two errors cancel corresponds to the probability of the event
\begin{equation*}
    D\cdot G\cdot B = G.
\end{equation*}
As the left-hand side is a Haar-random unitary, the probability of equality equals zero. 

An error for measurement gates corresponds to a situation where the measurement outcome differs from the actually measured state. 
As adaptive quantum circuits use measurements to control future quantum gates, a measurement error implies an incorrect control being used. 
Via a similar argument, the probability of an error undoing a measurement error is also zero. 

In the remainder, we determine the success probability of quantum circuits based on the success probability of the elementary operations used in the circuit. 
These elementary operations are single-qubit gates, two-qubit CNOT-gates, and measurements, as well as idling terms for other qubits during these operations. 
Additionally, we have the idling term during the conventional computations. 
\cref{tab:success_probabilities} summarizes these terms and their meaning. 
\begin{table}[th]
    \centering
    \caption{The considered success probabilities when analyzing state preparation protocols in the remainder of this chapter.}
    \begin{tabular}{c|l}
        Success term & Probability that ... \\
        \hline
        $p_s$ & a single-qubit gate succeeds \\
        $p_d$ & a two-qubit gate succeeds \\
        $p_m$ & a measurement returns the correct value \\
        $p_c$ & the intermediate computation returns the correct value \\
        $p_{ix}$ & a qubit remains coherent, while idling during an $x$-operation
    \end{tabular}
    \label{tab:success_probabilities}
\end{table}
We explicitly assume that intermediate conventional computations always return the correct answer, hence $p_c=1$. 
Note that $p_d$ relates to two qubits, whereas $p_{id}$ concerns individual qubits.

Most quantum devices require a decomposition of multi-qubit gates into single-qubit gate and CNOT-gates, supporting the choice for these error terms. 

In the remainder, we will use the term $P(X)$ to denote the success probability of a subroutine $X$. 
Similarly, $P(iX)$ denotes the probability that a qubit remains coherent while idling during the execution of $X$.
Additionally, $P(X, \textit{adapative})$ indicates the success probability of subroutine $X$, using an adaptive implementation.

As an example, the success probability of controlled-$U$-gates for some single-qubit gate~$U$ is given by 
\begin{equation}\label{eq:success_controlled_U}
    P(cU) = p_{s}^{3}p_{is}^{3}p_{d}^{2}.
\end{equation}
Here we use a standard decomposition of controlled-$U$-gates, see also~\cite[Corollary~4.2]{NielsenChuang:2010}.

The remainder of this work derives success probabilities for preparing the GHZ state and the $W$ state and provides first-order estimates of the relative theoretical performance of these protocols. 
This first-order estimate requires some assumptions about the seven variables used. 

We distinguish between easy operations, where the probability of an error is low, and hard operations, where the probability of an error is higher. 
This distinction applies to both the operations themselves and the corresponding idling times.
In most hardware realizations, single-qubit gate errors are significantly lower than two-qubit gate errors. 
Similarly, the gate times for single-qubit gates are significantly shorter than those for two-qubit gates. 
We therefore assume that single-qubit gates and idling during single-qubit gates are easy operations, while other operations are considered hard.
For a first-order estimate, we therefore assume $p_{s}\approx 1 \approx p_{is}$ and $p_{d} \approx p_{m}$, and $p_{id} \approx p_{im} \approx p_{ic}$. 
The assumptions are supported by observed success probabilities for quantum hardware, such as the error rates reported by IBM~\cite{IBMQuantumExperience:2024}.

When comparing protocols, we use approximate inequality signs ($\gtrsim$ and $\lesssim$) to indicate that we applied these assumptions to the success probabilities.

\section{Error analysis for GHZ state preparation}\label{sec:GHZ}
In this section, we derive an expression for the success probabilities of preparing a GHZ state using different non-adaptive and adaptive approaches. 
We consider non-adaptive approaches using a (restricted) all-to-all connectivity and a linear nearest-neighbor connectivity, and an adaptive approach using a linear nearest neighbor connectivity.
We also derive the success probability for a hybrid version that combines both approaches. 
These hardware connectivities cover most quantum hardware realizations. 

After deriving the success probabilities for the different approaches, we compare them to determine which protocol performs best in terms of the success probabilities of the individual parts of the circuit. 

This section derives the success probabilities for the different protocols and then compares them. 
The following theorem summarizes the results. 
\begin{theorem}\label{thm:error:GHZ}
    Let $\eps>0$ be a constant. 
    If $p_{d} \gtrsim (1+\eps)p_{id}^{\frac{n}{n-1}(\ceil{\log_2 n}/2 - 2)}$, then 
    \begin{equation*}
        P(GHZ_n, \textit{adaptive}) \gtrsim (1+\eps)^{2(n-1)} P(GHZ_n, \textit{all-to-all}).
    \end{equation*}
    If $p_{d} \gtrsim (1+\eps)p_{id}^{\frac{n}{n-1}(\ceil{n/2}/2 - 2)}$, then 
    \begin{equation*}
        P(GHZ_n, \textit{adaptive}) \gtrsim (1+\eps)^{2(n-1)} P(GHZ_n, \textit{linear}).
    \end{equation*}
\end{theorem}

Informally, the adaptive approach performs exponentially better than the non-adaptive approach using an all-to-all connectivity if $p_{d} > p_{id}^{\Omega(\log n)}$. 
This means that the adaptive approach performs exponentially better if the probability of a CNOT-gate introducing an error is at most the probability that an error is introduced while a qubit is idling for $\Omega(\log n)$ CNOT-gates. 
For the non-adaptive approach using a linear nearest-neighbor connectivity, we find that the adaptive approach performs exponentially better if $p_{d} > p_{id}^{\Omega(n)}$.
Both results are in line with what one might expect based on the circuit sizes of the different approaches.

\subsection{Success probability of GHZ state preparation}
Below we consider four possible approaches to prepare GHZ states and for each of them determine the success probability.
The four approaches considered are: 
\begin{enumerate}
    \item A non-adaptive approach using an all-to-all connectivity;
    \item A non-adaptive approach using a linear nearest-neighbor connectivity;
    \item An adaptive approach;
    \item A hybrid version of a non-adaptive and adaptive approach.
\end{enumerate}
We derive success probability expressions for preparing the GHZ state given by
\begin{equation*}
    \frac{1}{\sqrt{2}}\left(\ket{0}^{\otimes n}+\ket{1}^{\otimes n}\right).
\end{equation*}

\subsubsection{Non-adaptive approach using an all-to-all connectivity}
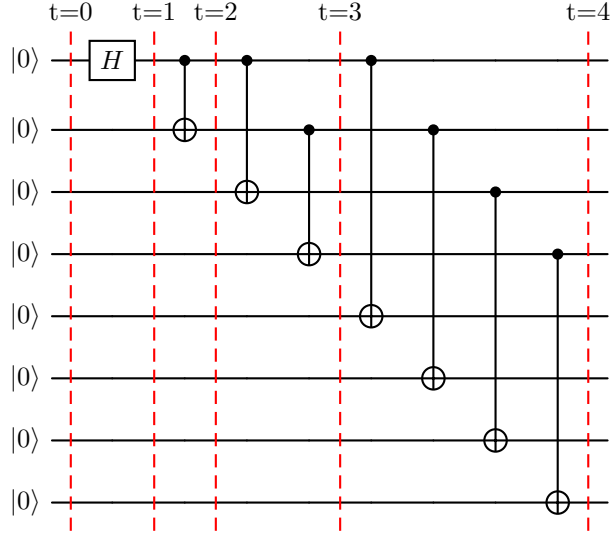
\begin{figure}[th]
    \centering
    \begin{quantikz}
    \lstick{\ket{0}}\slice{t=0} & \gate{H}\slice{t=1} & \ctrl{1}\slice{t=2} & \ctrl{2} & \slice{t=3} & \ctrl{4} & & & \slice{t=4} & \\
    \lstick{\ket{0}} & & \targ{} & & \ctrl{2} & & \ctrl{4} & & & \\
    \lstick{\ket{0}} & & & \targ{} & & & & \ctrl{4} & & \\
    \lstick{\ket{0}} & & & & \targ{} & & & & \ctrl{4} & \\
    \lstick{\ket{0}} & & & & & \targ{} & & & & \\
    \lstick{\ket{0}} & & & & & & \targ{} & & & \\
    \lstick{\ket{0}} & & & & & & & \targ{} & & \\
    \lstick{\ket{0}} & & & & & & & & \targ{} & 
    \end{quantikz}
    \caption{Circuit for preparing the GHZ state using an all-to-all connectivity for $n=8$. 
    The dotted lines indicate time steps and which gates can be applied in parallel.}
    \label{fig:q_circuit:GHZ_prep:all}
\end{figure}

\cref{fig:q_circuit:GHZ_prep:all} shows a quantum circuit to prepare the GHZ state for $n=8$ using an all-to-all connectivity.
In every subsequent time step, twice as many qubits can be targeted.

Let $k=\floor{\log_2 n}$ and let $m = n-2^k$. 
In the first layer, only a single one-qubit gate is applied. 
Afterwards, at time step $t=i$, $2^{i-1}$ two-qubit gates are applied, while all other qubits remain idle. 
After time step $t=k$, a total of $2^{k}-1$ qubits have been targeted by a CNOT-gate.
If $n$ is a power of two, the state has now been prepared. 
Otherwise, a single extra layer is necessary with $m$ CNOT-gates. 

For the success probability $P(GHZ_{n},\textit{all-to-all})$, we then arrive at the expression: 
\begin{align}
    & P(GHZ_{n},\textit{all-to-all}) \nonumber \\
    & \quad = p_s p_{is}^{n-1} \left(\Pi_{t=1}^{k} p_d^{2^{t-1}}p_{id}^{n-2^{t}}\right)p_d^{m}p_{id}^{(n-2m)(\ceil{\log_2 n}-k)} \nonumber \\
    & \quad = p_s p_{is}^{n-1}p_d^{n-1}p_{id}^{nk - 2^{k+1} + 2 + n(\ceil{\log_2 n}-\floor{\log_2 n}) -2m(\ceil{\log_2 n}-\floor{\log_2 n})} \nonumber \\
    & \quad = p_s p_{is}^{n-1}p_d^{n-1}p_{id}^{n\ceil{\log_2 n} - 2(n-m) + 2 -2m(\ceil{\log_2 n}-\floor{\log_2 n})} \nonumber \\
    & \quad = p_{s} p_{is}^{n-1} p_{d}^{n-1} p_{id}^{n(\ceil{\log_2 n}-2) + 2}. \label{eq:GHZ:P_all}
\end{align}
As $m(\ceil{\log_2 n}-\floor{\log_2 n}-1) = 0$, most terms in the exponent of $p_{id}$ cancel. 

To determine the probability that a qubit remains coherent while a GHZ state on $n$ qubits is prepared, we note that the circuit uses a single-qubit gate and then $\ceil{\log_2 n}$ layers of CNOT-gates. 
Combined, we have the success probability for idling qubits of 
\begin{equation*}
    P(iGHZ_{n},\textit{all-to-all}) = p_{is} p_{id}^{\ceil{\log_2 n}}.
\end{equation*}

\subsubsection{Non-adaptive approach using a linear nearest-neighbor connectivity}
\begin{figure}[th]
    \centering
    \begin{quantikz}
    \lstick{\ket{0}}\slice{t=0} & \slice{t=1} & \slice{t=2} & \slice{t=3} & \targ{}\slice{t=4} & \\
    \lstick{\ket{0}} & & & \targ{} & \ctrl{-1} & \\
    \lstick{\ket{0}} & \gate{H} & \ctrl{1} & \ctrl{-1} & & \\
    \lstick{\ket{0}} & & \targ{} & \ctrl{1} & & \\
    \lstick{\ket{0}} & & & \targ{} & \ctrl{1} & \\
    \lstick{\ket{0}} & & & & \targ{} &
    \end{quantikz}
    \caption{Circuit for preparing the GHZ state using a linear nearest-neighbor connectivity for $n=6$.
    The dotted lines indicate time steps and which gates can be applied in parallel.}
    \label{fig:q_circuit:GHZ_prep:linear}
\end{figure}
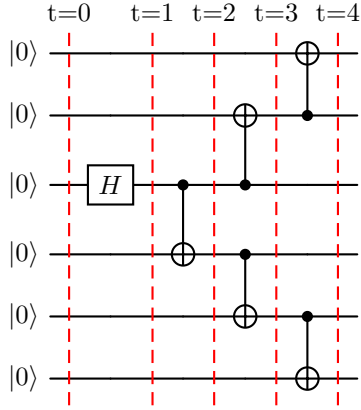

\cref{fig:q_circuit:GHZ_prep:linear} shows a quantum circuit to prepare the GHZ state for $n=6$ using a linear nearest-neighbor connectivity.
The key difference from the previous approach is that qubits can now only interact with their direct neighbors, and hence at most two CNOT-gates per layer can be applied. 

The circuit consists of the following steps: First, apply a Hadamard gate on qubit $\floor{\frac{n+1}{2}}$ and a CNOT-gate from that qubit to its direct neighbor with a higher index.
Let $k=\floor{\frac{n}{2}}$ and note that we can now have $k-1$ time steps consisting of $2$ CNOT-gates each. 
For odd $n$, we require a final layer consisting of a single CNOT-gate to include the last qubit in the GHZ state. 
We multiply the term in the exponent corresponding to the last layer with $n-2k$, since then the term vanishes for even $n$.

The overall probability of correctness $P(GHZ_{n},linear)$ is then given by
\begin{align}
    P(GHZ_{n},linear) & = p_{s} p_{is}^{n-1} p_{d}p_{id}^{n-2}\left(p_d^{2}p_{id}^{n-4}\right)^{k-1}\left(p_{d} p_{id}^{n-2}\right)^{n-2k} \nonumber \\
    & = p_{s} p_{is}^{n-1} p_{d}^{n-1} p_{id}^{n(\ceil{n/2}-2)+2}. \label{eq:GHZ:P_linear}
\end{align}
Note that for $n\le 6$, the performance using an all-to-all connectivity and a linear nearest-neighbor connectivity is the same. 
For $n\ge 7$, the approach using an all-to-all connectivity has higher success probability as it has fewer idling qubits. 

The idling term looks similar for this approach.
Only the exponent of $p_{id}$ differs, corresponding to the extra layers of CNOT-gates applied: 
\begin{equation*}
    P(iGHZ_{n},\textit{linear}) = p_{is} p_{id}^{\ceil{n/2}}.
\end{equation*}

Note that with little work, the derived expressions generalize to situations where a constant number of parallel CNOT-gates can be applied. 
In those cases, the exponent of $p_{id}$ in \cref{eq:GHZ:P_linear} will have a different constant, but will still have the same $\bigo(n^2)$ scaling.
This setting relates, for instance, to trapped-ion quantum computers that have an all-to-all connectivity, yet can only perform a constant number of parallel CNOT-gates~\cite{Moses:2023}.

\subsubsection{Adaptive approach}
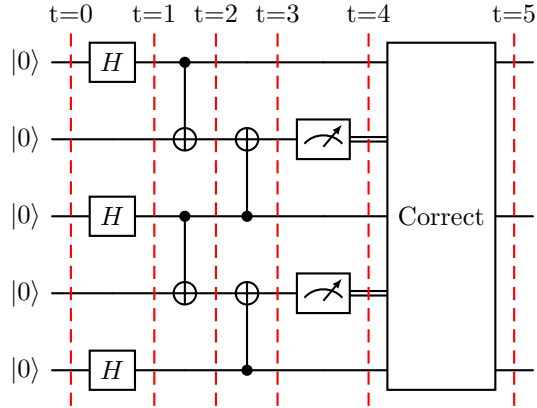
\begin{figure}[th]
    \centering
    \begin{quantikz}
    \lstick{\ket{0}}\slice{t=0} & \gate{H}\slice{t=1} & \ctrl{1}\slice{t=2} & \slice{t=3} & \slice{t=4} & \gate[5]{\text{Correct}}\slice{t=5} &  \\
    \lstick{\ket{0}} & & \targ{} & \targ{} & \meter{} & \setwiretype{c} \\
    \lstick{\ket{0}} & \gate{H} & \ctrl{1} & \ctrl{-1} & & &  \\
    \lstick{\ket{0}} & & \targ{} & \targ{} & \meter{} & \setwiretype{c} \\
    \lstick{\ket{0}} & \gate{H} & & \ctrl{-1} & & & 
    \end{quantikz}
    \caption{GHZ state preparation using an adaptive quantum circuit.}
    \label{fig:q_circuit:GHZ_prep:adaptive}
\end{figure}

Following the work of \citeauthor{Buhrman:2024}, we have an adaptive quantum circuit to prepare the GHZ state. 
\cref{fig:q_circuit:GHZ_prep:adaptive} shows the corresponding circuit for $n=3$.

The circuit uses $2n-1$ qubits and the circuit depth remains constant, even as $n$ grows. 
The success probability at time $t=4$ is given by 
\begin{equation*}
    p_s^n p_{is}^{n-1}p_d^{2(n-1)}p_{id}^2 p_m^{n-1}p_{im}^n.
\end{equation*}
A prefix sum computation on the measurement results indicates which qubits need correction. 
In the worst case, half of the qubits require a Pauli-$X$ correction. 
Assuming $p_s \le p_{is}$, which is valid as letting a qubit remain idle is generally better than manipulating it, we can lower bound the success probability:
\begin{equation}
    P(GHZ_{n}, \textit{adaptive}) \ge p_{s}^{n + \floor{n/2}} p_{is}^{n+\ceil{n/2}-1} p_{d}^{2(n-1)} p_{id}^{2} p_{m}^{n-1} p_{im}^{n} p_{ic}^{n}. \label{eq:GHZ:P_adaptive}
\end{equation}

An the corresponding idling term is given by
\begin{equation*}
    P(iGHZ_{n}, adaptive) \ge p_{is}^{2} p_{id}^{2} p_{im} p_{ic}.
\end{equation*}

\subsubsection{Hybrid approach}
We now combine the different state preparation approaches: 
First, use a standard approach to prepare $k$ small GHZ states.
Then, use an adaptive approach to join these small GHZ states together.

We assume that $k$ perfectly divides $n$, such that $n=kg$ for some positive integer~$g$. 
The situation where we prepare GHZ states of different sizes and then recombine them follows via a similar derivation. 
This gives $k$ terms for non-adaptively preparing a GHZ state of size $g$ (\cref{eq:GHZ:P_all,eq:GHZ:P_linear}), with $k$ corresponding idling terms, multiplied by an adaptive term for preparing a GHZ state of size $k$ (\cref{eq:GHZ:P_adaptive}) and $n-k$ corresponding idling terms. 

Filling in the terms derived above, we obtain the following two expressions for the success probability, depending on the used connectivity:
\begin{align}
    & P(GHZ_{n,k},\textit{hybrid-all}) \ge \nonumber \\
    & \qquad p_{s}^{2k + \floor{n/2}} p_{is}^{3n - k + \ceil{n/2} - 1} p_{d}^{n + k - 2} p_{id}^{(n + k)\ceil{\log_2 g} + 2} p_{m}^{k - 1} p_{im}^{n} p_{ic}^{n}. \label{eq:GHZ:P_hybrid_all} \\
    & P(GHZ_{n,k},\textit{hybrid-linear}) \ge \nonumber \\
    & \qquad p_{s}^{2k + \floor{n/2}} p_{is}^{3n - k + \ceil{n/2} - 1} p_{d}^{n + k - 2} p_{id}^{(n+k)\ceil{g/2} + 2} p_{m}^{k - 1} p_{im}^{n} p_{ic}^{n}. \label{eq:GHZ:P_hybrid_linear}
\end{align}

\subsection{Comparing GHZ state preparation approaches}
This section provides a comparison between the state preparation approaches in terms of the success probabilities in order to prove \cref{thm:error:GHZ}.
In the proof of the theorem, we use $\gtrsim$ to stress that we used assumptions on the relative magnitude of the different success probabilities, see also \cref{sec:error_model}. 
The theorem follows directly from these comparisons.

\subsubsection{Non-adaptive all-to-all versus adaptive}\label{sec:GHZ:error_all_adaptive}
Comparing \cref{eq:GHZ:P_adaptive,eq:GHZ:P_all}, we see that the inequality of the success probability holds precisely if
\begin{equation*}
    p_{s}^{n + \floor{n/2} - 1} p_{is}^{\ceil{n/2}} p_{d}^{n-1} p_{m}^{n-1} p_{im}^{n} p_{ic}^{n} \ge p_{id}^{n(\ceil{\log_2 n}-2)}.
\end{equation*}
Applying the assumptions on the success probabilities discussed in \cref{sec:error_model}, we see that this expression reduces to 
\begin{equation*}
    p_{d}^{2(n-1)} \gtrsim p_{id}^{n(\ceil{\log_2 n} - 4)}.
\end{equation*}
Hence, the adaptive approach performs better if 
\begin{equation}\label{eq:error:GHZ_all_adaptive}
    p_{d} \gtrsim p_{id}^{\frac{n}{n-1}(\ceil{\log_2 n}/2 - 2)}.
\end{equation} 
That is, the adaptive approach outperforms the standard approach using an all-to-all connectivity if the probability of a CNOT-gate introducing an error is at most the probability of a qubit idling during $\frac{n}{n-1}(\ceil{\log_2 n}/2 - 2)$ CNOT-gates picking up an error.

Now note that we can rewrite \cref{eq:GHZ:P_adaptive} in terms of \cref{eq:GHZ:P_all}, using the assumptions on the success probabilities, as
\begin{equation}
    P(GHZ_n,\textit{adaptive}) \gtrsim p_{d}^{2(n-1)} p_{id}^{n(4-\ceil{\log_2 n})} P(GHZ_n,\textit{all}).
\end{equation}
Hence, for $p_{d} = (1+\eps)p_{id}^{\frac{n}{n-1}(\ceil{\log_2 n}/2 - 2)}$ we obtain
\begin{equation}
    P(GHZ_n,\textit{adaptive}) \gtrsim (1+\eps)^{2(n-1)} P(GHZ_n,\textit{all}),
\end{equation}
proving the first statement of \cref{thm:error:GHZ}.

Note that for large $n$, \cref{eq:error:GHZ_all_adaptive} simplifies to
\begin{equation} \label{eq:GHZ:error_informal_all}
    p_{d} \gtrsim p_{id}^{\Omega(\log_2 n)},
\end{equation}
corresponding with what we expected based on the circuit size.

\subsubsection{Non-adaptive linear nearest-neighbor versus adaptive}\label{sec:GHZ:error_linear_adaptive}
Comparing \cref{eq:GHZ:P_adaptive,eq:GHZ:P_linear}, we see that the inequality of the success probability holds precisely if
\begin{equation*}
    p_{s}^{n + \floor{n/2} - 1} p_{is}^{\ceil{n/2}} p_{d}^{n-1} p_{m}^{n-1} p_{im}^{n} p_{ic}^{n} \ge p_{id}^{n(\ceil{n/2}-2)}.
\end{equation*}
Applying the assumptions on the success probabilities, we see that this expression reduces to 
\begin{equation*}
    p_{d}^{2(n-1)} \gtrsim p_{id}^{n(\ceil{n/2}-4)}.
\end{equation*}
Hence, the adaptive approach performs better if 
\begin{equation}\label{eq:error:GHZ_linear_adaptive}
    p_{d} \gtrsim p_{id}^{\frac{n}{n-1}(\ceil{n/2}/2 - 2)}.
\end{equation} 
That is, the adaptive approach outperforms the standard protocol using a linear nearest-neighbor connectivity if the probability of a CNOT-gate introducing an error is at most the probability of a qubit idling during $\frac{n}{n-1}(\ceil{n/2}/2 - 2)$ CNOT-gates picking up an error.

Now note that we can rewrite \cref{eq:GHZ:P_adaptive} in terms of \cref{eq:GHZ:P_linear}, using the assumptions on the success probabilities, as
\begin{equation}
    P(GHZ_n,\textit{adaptive}) \gtrsim p_{d}^{2(n-1)} p_{id}^{n(4-\ceil{n/2})} P(GHZ_n,\textit{linear}).
\end{equation}
Hence, for $p_{d} = (1+\eps)p_{id}^{\frac{n}{n-1}(\ceil{n/2}/2 - 2)}$ we obtain
\begin{equation}
    P(GHZ_n,\textit{adaptive}) \gtrsim (1+\eps)^{2(n-1)} P(GHZ_n,\textit{linear}),
\end{equation}
proving the second statement of \cref{thm:error:GHZ}.

Note that for large $n$, \cref{eq:error:GHZ_linear_adaptive} simplifies to
\begin{equation} \label{eq:GHZ:error_informal_linear}
    p_{d} \gtrsim p_{id}^{\Omega(n)},
\end{equation}
corresponding with what we expected based on the circuit size. 
Settings where a constant number of parallel gates is allowed, without restrictions on the topology, will have a similar asymptotic scaling.

\subsubsection{Comparison with the hybrid approach}
We now compare the two standard approaches with their corresponding hybrid version. 
In the hybrid approach, $k$ GHZ states of $g$ qubits each are combined using the adaptive approach to prepare a GHZ state on $n=kg$ qubits. 
Therefore, we express the relative performance of the approaches in terms of $k$ and~$g$. 

For the hybrid approach using an all-to-all connectivity, we find that it outperforms the standard approach using the same connectivity if
\begin{equation*}
    P(GHZ_{n,k},\textit{hybrid-all}) \ge P(GHZ_{n}, \textit{all}).
\end{equation*}
Using \cref{eq:GHZ:P_hybrid_all,eq:GHZ:P_all}, this expression simplifies to
\begin{equation*}
    p_{s}^{2k + \floor{n/2} - 1} p_{is}^{2n - k + \ceil{n/2}} p_{d}^{k - 1} p_{m}^{k - 1} p_{im}^{n} p_{ic}^{n} \ge p_{id}^{n(\ceil{\log_2 n} - \ceil{\log_2 g} - 2) - k\ceil{\log_2 g}}.
\end{equation*}
Again applying the assumptions on the success probabilities, we see that this expression reduces to
\begin{equation} \label{eq:GHZ:error_pd_id_all}
    p_{d} \gtrsim p_{id}^{\frac{n}{2(k-1)}(\ceil{\log_2 n} - \ceil{\log_2 g} - 4) - \frac{k}{2(k-1)}\ceil{\log_2 g}}.
\end{equation}
Let $\eps>0$ and $p_{d} = (1+\eps) p_{id}^{\frac{n}{2(k-1)}(\ceil{\log_2 n} - \ceil{\log_2 g} - 4) - \frac{k}{2(k-1)}\ceil{\log_2 g}}$, then 
\begin{equation}
    P(GHZ_{n,k},\textit{hybrid-all}) \ge (1+\eps)^{2(k-1)} P(GHZ_n,\textit{all}).
\end{equation}
Hence, the hybrid approach performs exponentially better than the standard approach. 

For the hybrid approach using a linear nearest-neighbor connectivity, we find that it outperforms the standard approach using the same connectivity if
\begin{equation*}
    P(GHZ_{n,k},\textit{hybrid-linear}) \ge P(GHZ_{n}, \textit{linear}).
\end{equation*}
Using \cref{eq:GHZ:P_hybrid_linear,eq:GHZ:P_linear}, this expression simplifies to
\begin{equation*}
    p_{s}^{2k + \floor{n/2} - 1} p_{is}^{2n - k + \ceil{n/2}} p_{d}^{k - 1} p_{m}^{k - 1} p_{im}^{n} p_{ic}^{n} \ge p_{id}^{n(\ceil{n/2} - \ceil{g/2} - 2) - k\ceil{g/2}}.
\end{equation*}
Again applying the assumptions on the success probabilities, we see that this expression reduces to
\begin{equation} \label{eq:GHZ:error_pd_id_linear}
    p_{d} \gtrsim p_{id}^{\frac{n}{2(k-1)}(\ceil{n/2} - \ceil{g/2} - 4) - \frac{k}{2(k-1)}\ceil{g/2}}.
\end{equation}
Let $\eps>0$ and $p_{d} = (1+\eps) p_{id}^{\frac{n}{2(k-1)}(\ceil{n/2} - \ceil{g/2} - 4) - \frac{k}{2(k-1)}\ceil{g/2}}$, then 
\begin{equation}
    P(GHZ_{n,k},\textit{hybrid-linear}) \ge (1+\eps)^{2(k-1)} P(GHZ_n,\textit{linear}).
\end{equation}
Hence, the hybrid approach performs exponentially better than the standard approach. 

For both hybrid approaches, we can obtain an informal estimate similar to \cref{eq:GHZ:error_informal_all,eq:GHZ:error_informal_linear}. 
Using that $n=kg$, $\ceil{x}\approx x \approx \floor{x}$ for any $x\in\R$, and $\frac{x}{x-1}\approx 1$ for large $x\in \R$, we see that \cref{eq:GHZ:error_pd_id_all,eq:GHZ:error_pd_id_linear} simplify to
\begin{equation}
    p_{d} \gtrsim p_{id}^{\Omega(g\log_2 k)}
\end{equation}
and
\begin{equation}
    p_{d} \gtrsim p_{id}^{\Omega(ng)},
\end{equation}
respectively.

Note that for $g=\bigo(1)$ (and hence $k=\Theta(n)$), the hybrid approaches perform similarly to the adaptive approach. 

\section{Implementation on quantum hardware}\label{sec:hardware_implementation}
In the previous sections, we compared three GHZ state generation approaches.
In this section, we implement the approaches on quantum hardware and compare the results with our theoretical results. 
We first discuss the nuances and practicalities of implementing a protocol on currently available quantum hardware. 
Then, we present the results for the selected quantum device. 
Our analysis used a worst-case error model, overestimating the errors and their effect in quantum hardware. 
We thus expect to see differences compared to the theoretical derivations.

\subsection{Setup and implementation details}
We implemented the protocols on the IBM Brisbane device, which is based on superconducting technology~\cite{IBMQuantumExperience:2024}. 
The device has 127 qubits. 
Given the device's topology and the qubit requirements of the adaptive approach, we can prepare a GHZ state on at most $n=55$ qubits using the non-adaptive linear nearest-neighbor approach and the adaptive approach. 
For each approach, we give the measurement outcomes as a quasi-probability distribution based on $4{,}096$ samples. 
We also provide the expected success probabilities based on \cref{eq:GHZ:P_linear,eq:GHZ:P_adaptive} and using the parameters of the device.

Note that the measurements used to obtain this quasi-probability distribution can introduce errors themselves. 
Furthermore, we cannot detect phase errors, as measurements are performed only in the Pauli-$Z$ basis. 
However, as both approaches apply $n$ measurements at the end of the circuit, we ignore the resulting decoherence, expecting its impact on both outcomes to be approximately the same.

We expect differences between the hardware results and the theoretical results for multiple reasons. 
First, the worst-case error model used likely gives an upper bound on errors in practice. 
Second, quantum hardware systems typically have a limited set of native gates, requiring that some gates used in an algorithm are decomposed into native gates. 
Third, conventional pre- and post-processing techniques can help reduce the circuit depth and the errors in the circuits. 

In a noiseless situation, we expect the circuits to give a near-uniform distribution between the all-zeros and all-ones outcomes upon measurement. 
In a noisy setting, we expect to often find the two correct measurement outcomes, but we also expect to find other measurement results due to implementation imperfections.

For the theoretical success probabilities, we use the success probabilities of the individual terms in a circuit. 
Most of these terms are provided by the quantum hardware provider in terms of error probabilities.
However, the idling terms for CNOT-gates and measurements are not provided and are derived manually from the corresponding gate times and the median $T_2$ decay time.

When implementing the adaptive circuit, we found that classical computations were restricted to control of future quantum gates only. 
No additional computations were possible. 
The resulting implementation thus is suboptimal, using significantly more gates than initially required. 

Finally, we set the optimization level of the IBM compiler to $1$, using only necessary optimizations for implementing the circuit and simple simplifications. 

We now give the code used to generate the results on quantum hardware. 
\begin{python}
from qiskit import QuantumCircuit, QuantumRegister, ClassicalRegister
from qiskit_ibm_runtime import QiskitRuntimeService, SamplerV2
from qiskit.transpiler.preset_passmanagers import generate_preset_pass_manager

API_token = "<your token here>"
backend_name = "ibm_brisbane"
n_qubits = 10
service = QiskitRuntimeService(channel="ibm_quantum", token=API_token)
backend = service.backend(backend_name)
pm = generate_preset_pass_manager(backend=backend, optimization_level=1)

# Adaptive circuit
qrm = QuantumRegister(n_qubits)
qrx = QuantumRegister(n_qubits - 1)
crx = ClassicalRegister(n_qubits-1, name="intermediate_result")
crm = ClassicalRegister(n_qubits, name="final_result")
qcircuit_adaptive = QuantumCircuit(qrm,qrx, crx, crm, name="GHZ")
for i in range(n_qubits):
    qcircuit_adaptive.h(qrm[i])
for i in range(n_qubits-1):
    qcircuit_adaptive.cx(qrm[i], qrx[i])
for i in range(1,n_qubits):
    qcircuit_adaptive.cx(qrm[i], qrx[i-1])

for i in range(n_qubits - 1):
    qcircuit_adaptive.measure(qrx[i], crx[i])
    
for i in range(n_qubits-1):
    with qcircuit_adaptive.if_test((crx[i], 1)):
        for j in range(i+1, n_qubits):
            qcircuit_adaptive.x(qrm[j])

qcircuit_adaptive.measure(qrm, crm)

isa_circuit_adaptive = pm.run(qcircuit_adaptive)
sampler = Sampler(backend)
job = sampler.run([(isa_circuit_adaptive)])
result_adaptive = job.result()

# Non-adaptive circuit
qrm = QuantumRegister(n_qubits)
crm = ClassicalRegister(n_qubits)

start_qubit = (n_qubits + 1) // 2 - 1
k = n_qubits//2
qcircuit_standard = QuantumCircuit(qrm, crm, name="GHZ")

qcircuit_standard.h(start_qubit)
qcircuit_standard.cx(start_qubit, start_qubit +1)
for i in range(k-1):
    qcircuit_standard.cx(start_qubit-i, start_qubit-i-1)
    qcircuit_standard.cx(start_qubit+1+i, start_qubit+2+i)

if n_qubits - 2*k: # Check if we need a final layer
    qcircuit_standard.cx(1, 0)
    
qcircuit_standard.measure(qrm, crm)

isa_circuit_standard = pm.run(qcircuit_standard)
sampler = Sampler(backend)
job = sampler.run([(isa_circuit_standard)])
result_standard = job.result()
\end{python}

We proceed by comparing the theoretical success probabilities from \cref{eq:GHZ:P_linear,eq:GHZ:P_adaptive} with the implementation results.
As mentioned, we expect these success probabilities to give a lower bound on the actual success probabilities. 
Additionally, we give the expected running time of each circuit based on the obtained gate and measurement times. 
Finally, we show the measurement results for different values of $n$ for both approaches, allowing us to compare practical performance and observe how the success probabilities change as $n$ grows.

\subsection{Implementation on the IBM Brisbane device}
This section presents the results of the hardware experiments run on the IBM Brisbane device. 
\cref{tab:success_probabilities_IBM_Brisbane} summarizes the success probabilities of different gates of this device, using a $T_2$ value of $131.71\mu$s.
\begin{table}[th]
    \centering
    \begin{tabular}{c|l|l}
        Success term & Value & Obtained via \\
        \hline
        $p_{s}$ & $1 - 2.530\cdot 10^{-4}$ & Provided by IBM \\
        $p_{is}$ & $1 - 2.530\cdot 10^{-4}$ & Provided by IBM \\
        $p_{d}$ & $1 - 9.442\cdot 10^{-3}$ & Provided by IBM \\
        $p_{id}$ & $1 - 4.998\cdot 10^{-3}$ & Based on a gate time of $660$ ns \\
        $p_{m}$ & $1-1.600\cdot 10^{-2}$ & Provided by IBM \\
        $p_{im}$ & $1-9.822 \cdot 10^{-3}$ & Based on a measurement time of $1300$ ns \\
        $p_{ic}$ & $1-9.822 \cdot 10^{-3}$ & Equal to $p_{im}$
    \end{tabular}
    \caption{Success probabilities of IBM Brisbane device, based on calibration on December 13, 2024 at 09.30 (UTC+2).}
    \label{tab:success_probabilities_IBM_Brisbane}
\end{table}

\cref{eq:GHZ:P_linear,eq:GHZ:P_adaptive} give the following two success probabilities for the largest GHZ state preparable on this device
\begin{align*}
    P(Brisbane, GHZ_{55}, \textit{linear}) & = p_{s} p_{is}^{54} p_{d}^{54} p_{id}^{1432} \\
    & = 4.52 \cdot 10^{-4} \\
    P(Brisbane, GHZ_{55}, \textit{adaptive}) & \ge p_{s}^{82} p_{is}^{83} p_{d}^{108} p_{id}^{2} p_{m}^{54} p_{im}^{55} p_{ic}^{55} \\
    & = 4.82 \cdot 10^{-2}.
\end{align*}
We see a factor $100$ difference in the theoretical success probabilities with a worst-case error model, favoring the adaptive approach. 

To determine the time duration of both approaches, we estimate the single-qubit gate time using $p_{s}$ and $T_2$ to be approximately $33$ ns. 
Using a two-qubit gate time of $660$ ns and a measurement time of $1300$ ns, we find running times
\begin{align*}
    T(Brisbane, GHZ_{55}, \textit{linear}) & = 33 \text{ns} + 28*660 \text{ns} = 18.51 \mu\text{s} \\
    T(Brisbane, GHZ_{55}, \textit{adaptive}) & = 2(33 + 660 + 1300) \text{ns} = 3.99 \mu\text{s}.
\end{align*}
Hence, we expect the adaptive approach to have a shorter running time.

Below, we give the results of the hardware implementations for different $n$.
For small $n$, all measurement outcomes are given. 
For larger $n$, we instead give aggregated results, grouping the results of strings with the same Hamming weight.

Based on the success probabilities shown in \cref{tab:success_probabilities_IBM_Brisbane} and the theoretical relation between $p_{d}$ and $p_{id}$ shown in \cref{thm:error:GHZ}, we expect the adaptive approach to outperform the standard approach for $n\ge 15$. 

\begin{figure}
    \centering
    \includegraphics[width=0.32\textwidth]{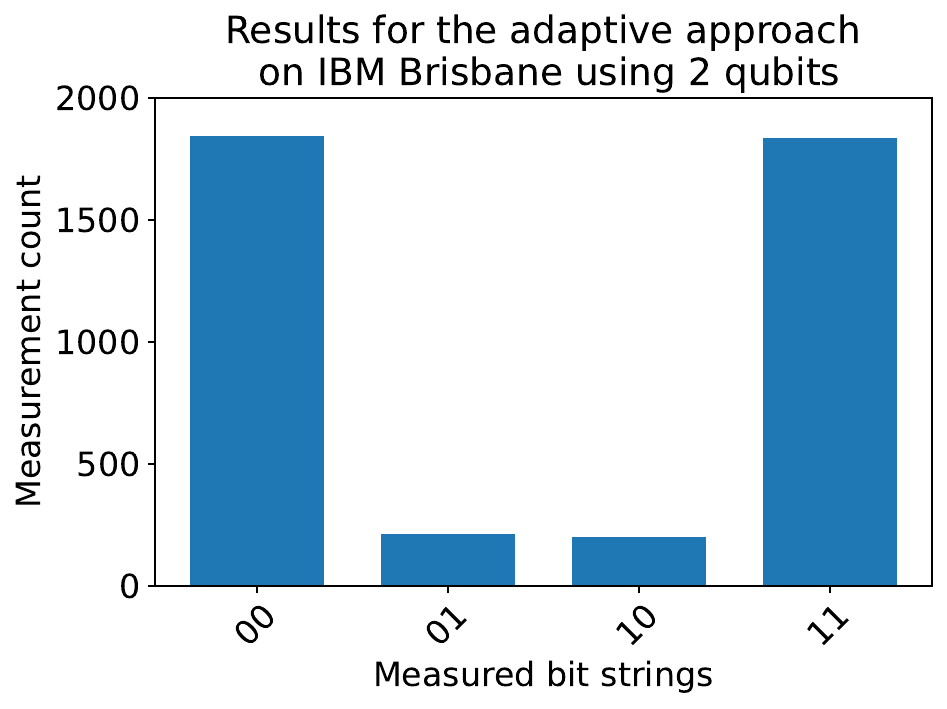}
    \includegraphics[width=0.32\textwidth]{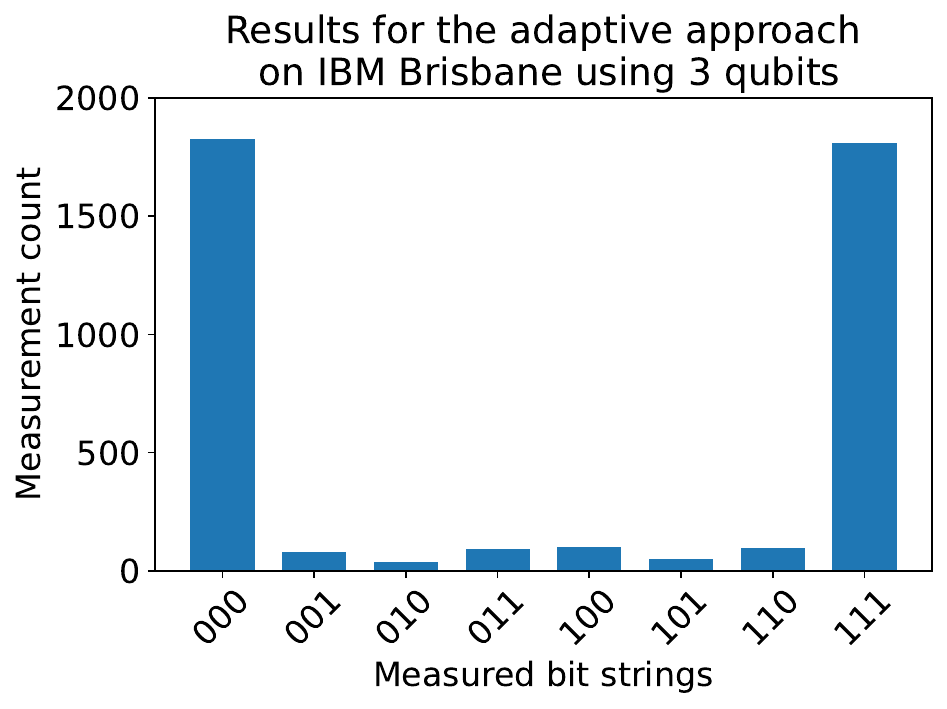}
    \includegraphics[width=0.32\textwidth]{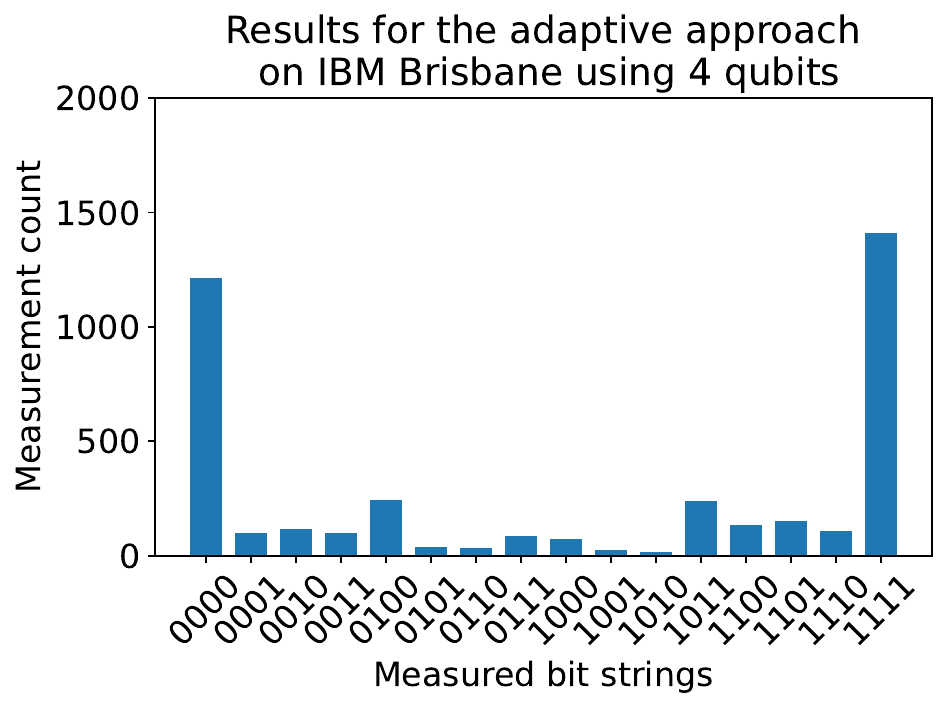}
    \includegraphics[width=0.32\textwidth]{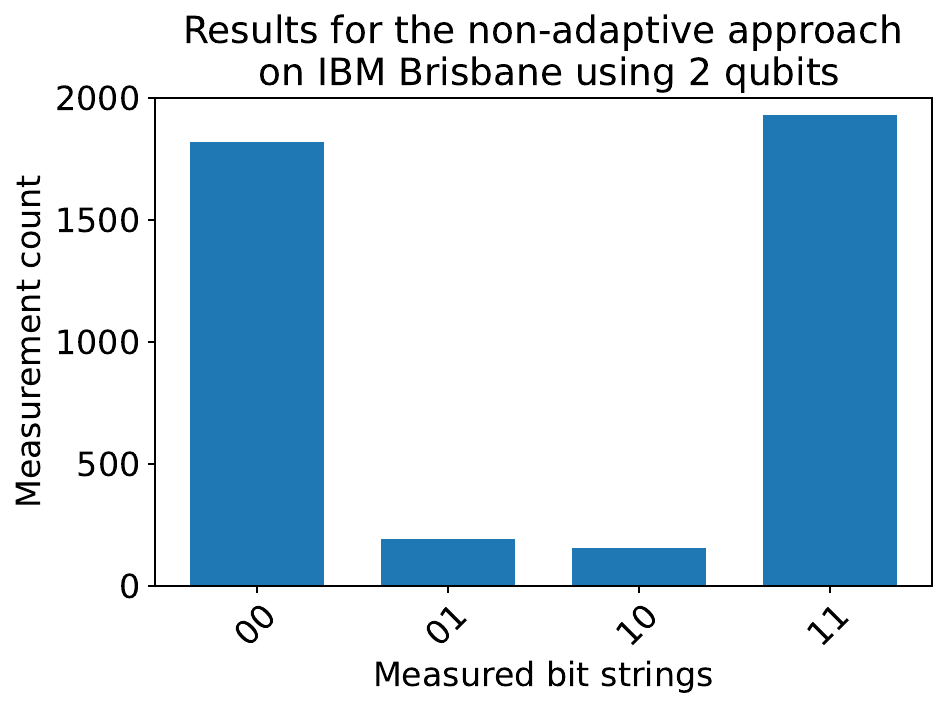}
    \includegraphics[width=0.32\textwidth]{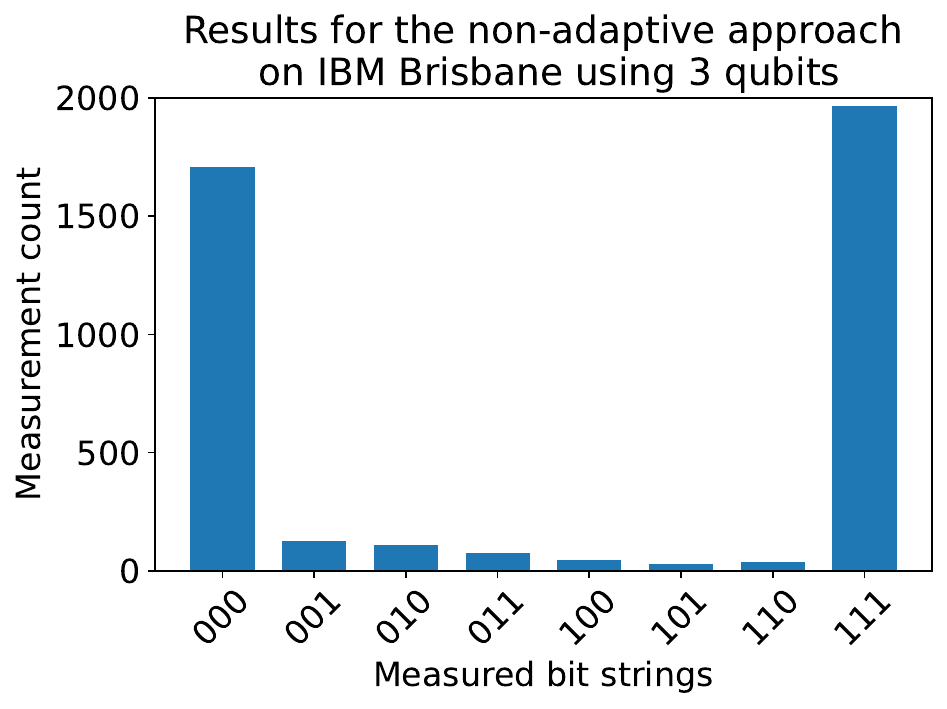}
    \includegraphics[width=0.32\textwidth]{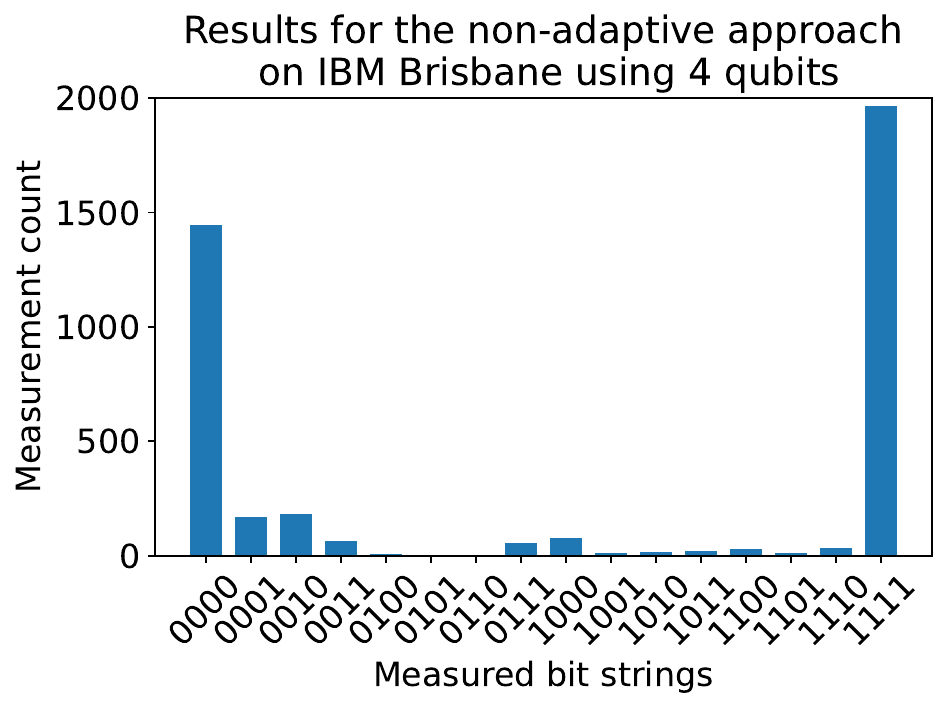}
    \caption{Measurement results for preparing a GHZ state on few qubits on the IBM Brisbane device using the adaptive approach and the standard approach.
    Horizontally, the different measurement results are shown and the height of the bars shows how often that measurement result is found.}
    \label{fig:error:Brisbane:adaptive_standard_small_n}
\end{figure}

\cref{fig:error:Brisbane:adaptive_standard_small_n} shows the results for small $n$. 
The standard approach slightly outperforms the adaptive approach, as expected based on the success probabilities. 

\begin{figure}
    \centering
    \includegraphics[width=0.45\textwidth]{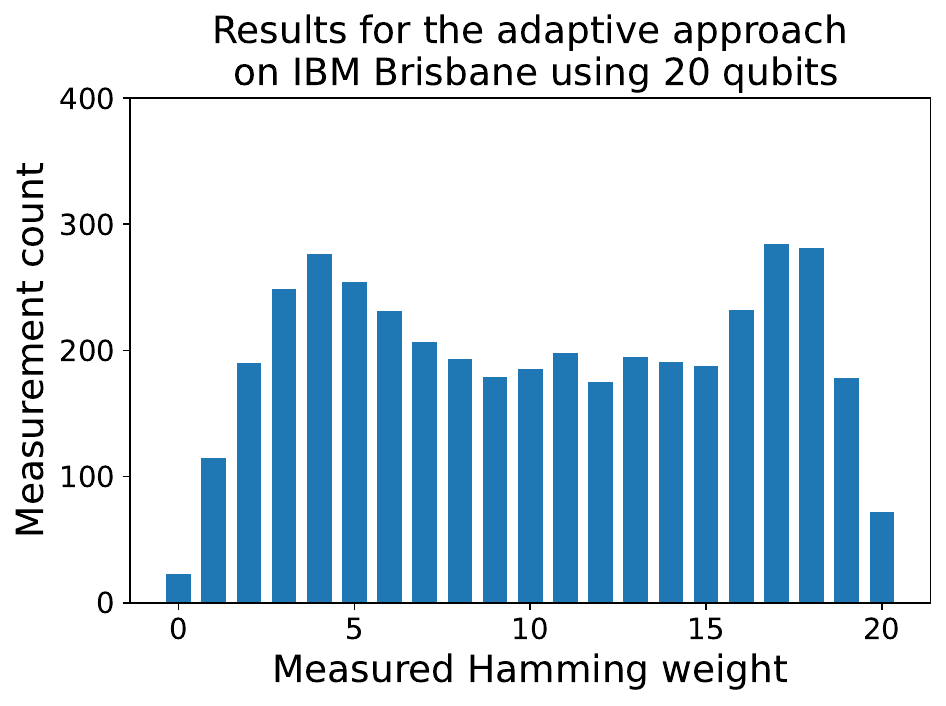}
    \includegraphics[width=0.45\textwidth]{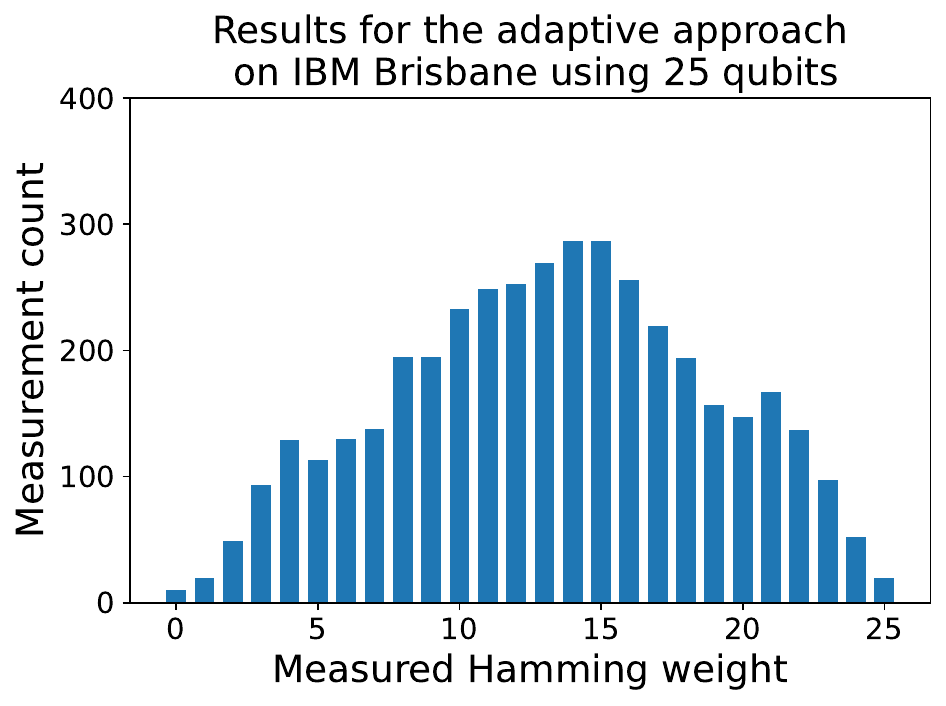}
    \includegraphics[width=0.45\textwidth]{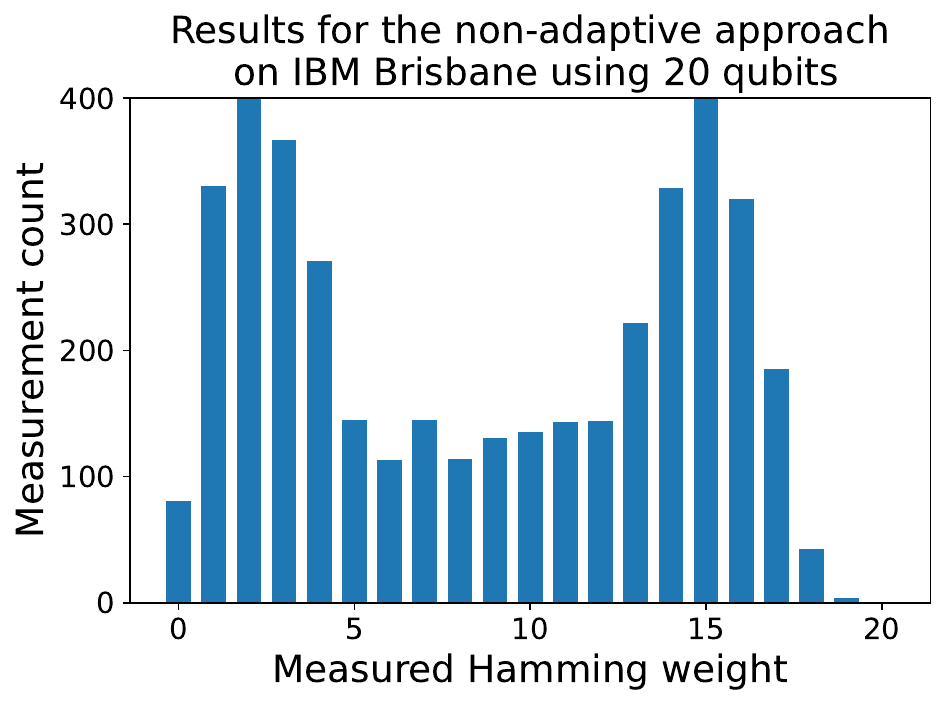}
    \includegraphics[width=0.45\textwidth]{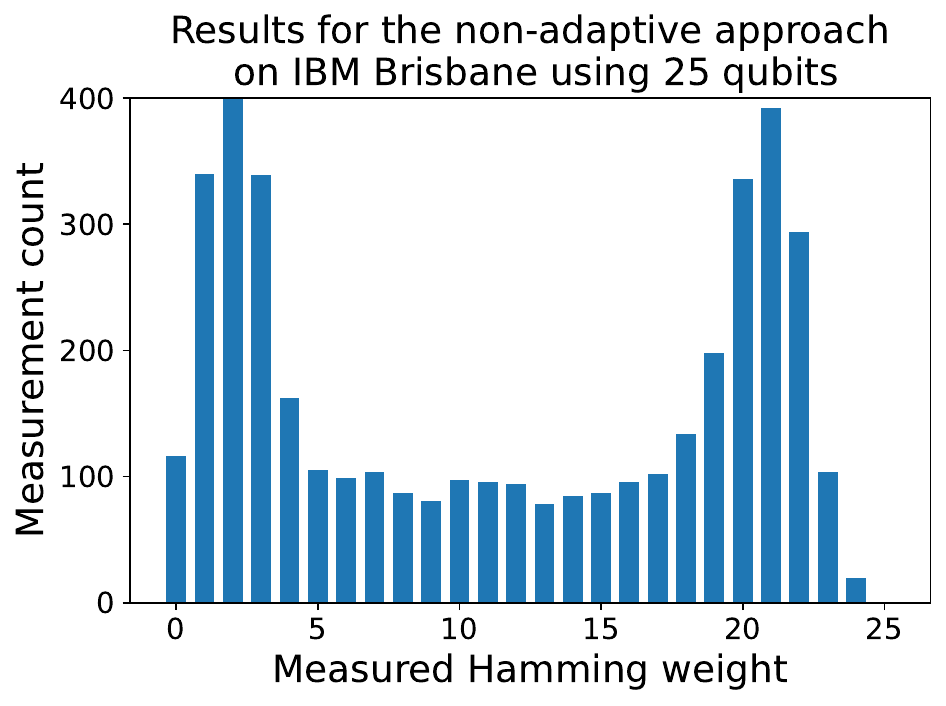}
    \caption{Measurements results for preparing a GHZ state on $n=20$ and $n=25$ qubits on the IBM Brisbane device using the adaptive approach and the standard approach. 
    Horizontally, the different measured Hamming weights are shown and the height of the bars shows how often that Hamming weight was found.}
    \label{fig:error:Brisbane:adaptive_standard_n20_25}
\end{figure}

\cref{fig:error:Brisbane:adaptive_standard_n20_25} shows the aggregated results for $n=20$ and $n=25$, where measurement outcomes with the same Hamming weight are grouped. 
The aggregated results show the magnitude of the errors. 
With the adaptive approach, both expected outcomes ($0$ and $n$) are measured, but the results are somewhat uniform.
In contrast, the standard approach did not return the all-ones string in any measurement, but does show two distinct peaks near the low and high Hamming weight outcomes. 

\begin{figure}
    \centering
    \includegraphics[width=0.3\textwidth]{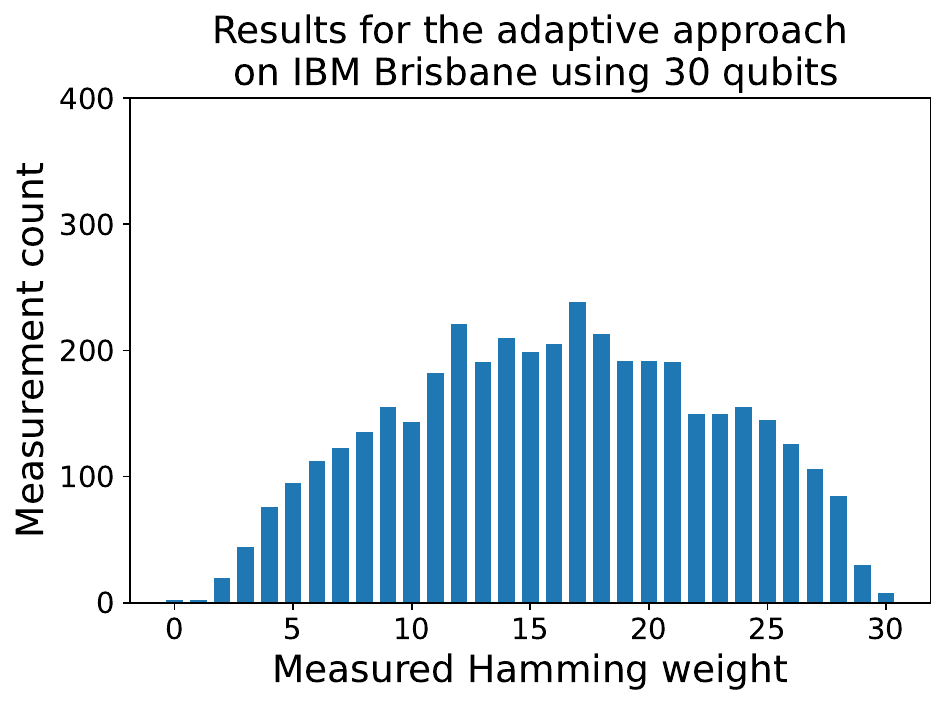}
    \includegraphics[width=0.3\textwidth]{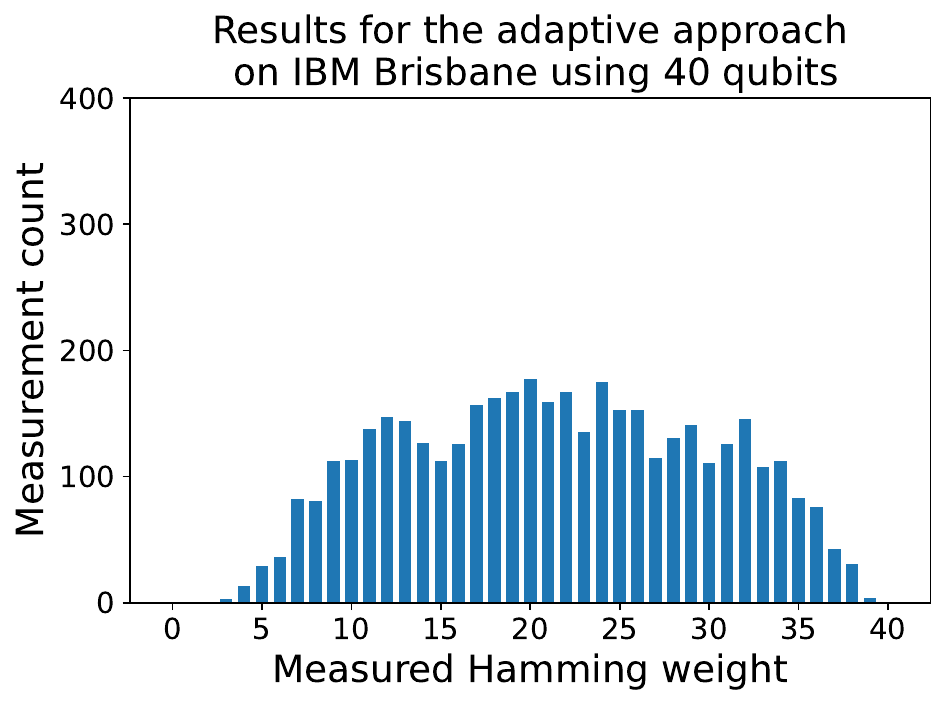}
    \includegraphics[width=0.3\textwidth]{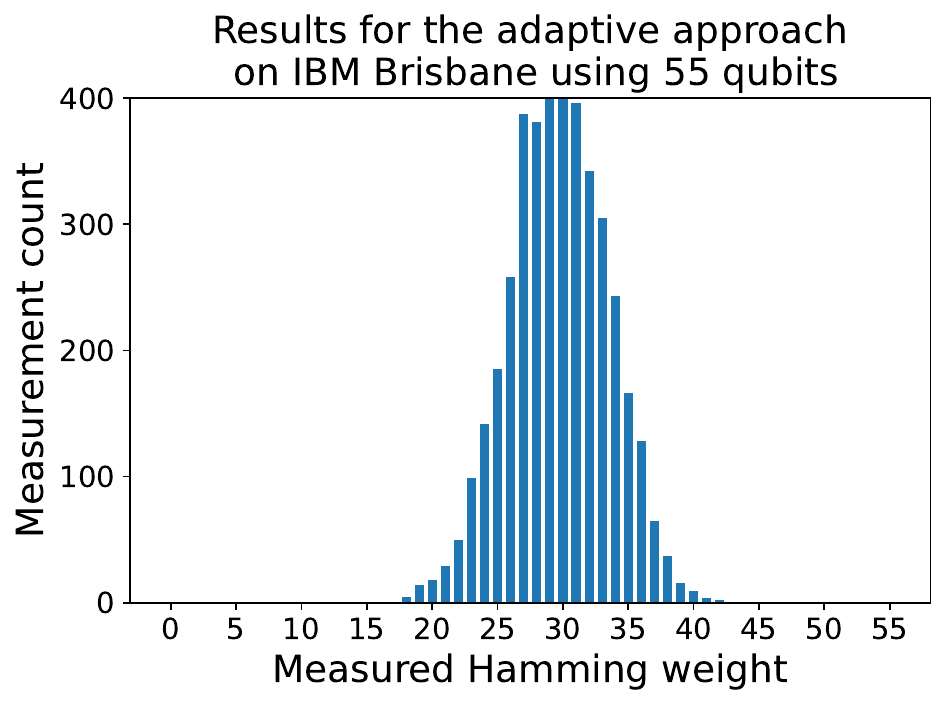}
    \includegraphics[width=0.3\textwidth]{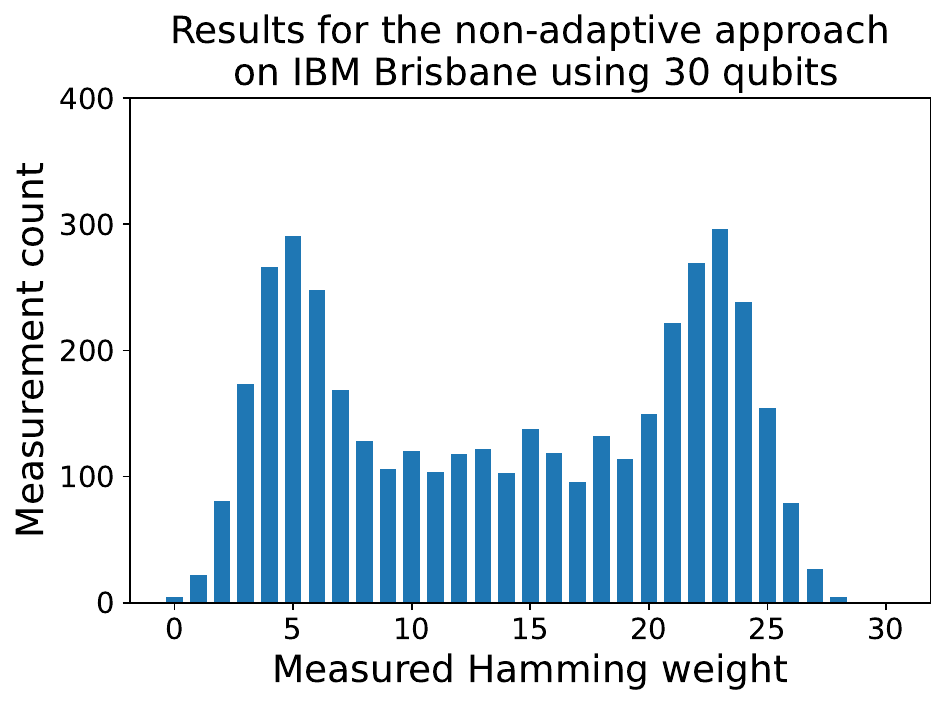}
    \includegraphics[width=0.3\textwidth]{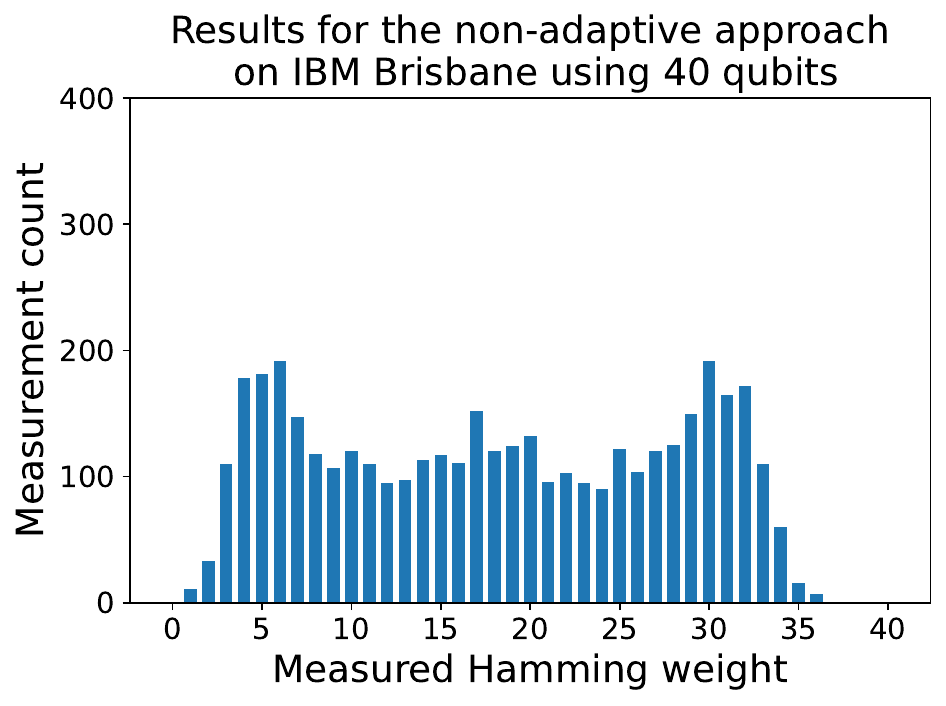}
    \includegraphics[width=0.3\textwidth]{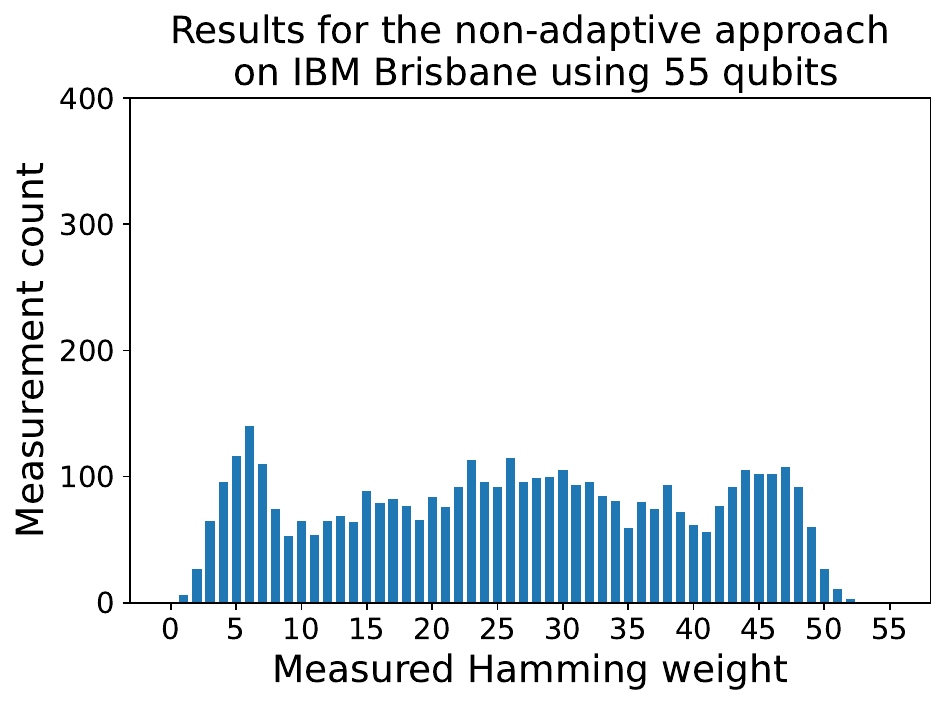}
    \caption{Measurement results for preparing a GHZ state on $n=30$, $n=40$, and $n=55$ qubits on the IBM Brisbane device using the adaptive approach and the standard approach.
    Horizontally, the different measured Hamming weights are shown and the height of the bars shows how often that Hamming weight was found.}
    \label{fig:error:Brisbane:adaptive_standard_large_n}
\end{figure}

\cref{fig:error:Brisbane:adaptive_standard_large_n} shows the results for large $n$. 
The adaptive approach appears to produce a normal distribution, as seen for $n=25$. 
This suggests that the adaptive approach samples from a uniform superposition over all bit strings. 
The results for the standard approach show similarities with the expected distribution, with a little more weight towards the two extremes of the distribution. 

For the adaptive approach, the limitation on the classical computations make that we have to additionally perform up to $54$ correction terms, instead of just one. 
This affects both the probability of successfully preparing the GHZ state, as well as the time duration of the entire quantum circuit.

\section{Error analysis for \texorpdfstring{$W$}{W}-state preparation}\label{sec:W_state}
In this section, we derive an expression for the success probability of preparing $W$-states with a standard approach using a linear nearest-neighbor connectivity and with the adaptive approach introduced in~\cite{Buhrman:2024}.
The standard approach has depth $\bigo(n)$ and uses $n$ qubits, whereas the adaptive approach uses $\bigo(n\log (n)\log\log (n))$ qubits and has constant depth. 
Based on the circuit sizes, our intuition suggests that the adaptive approach performs better if 
\begin{equation*}
    p_{d} \gtrsim p_{id}^{\Omega(n/(\log (n)\log\log (n)))}.
\end{equation*}
In the next sections, we will derive success probabilities for both approaches and will see that our intuition is indeed correct, as summarized in the next theorem. 
\begin{theorem}\label{thm:error:W_state}
    Let $n=2^k$ for some integer $k$ and let $\eps >0$ be a constant. 
    If $p_d \gtrsim (1+\eps)p_{id}^{3n/(59\log_2 (n) \log_2\log_2 (n))}$, then, with respect to the most significant terms, $P(W,\textit{adaptive}) \gtrsim (1+\eps)^{59n / (\log_2 (n)\log_2\log_2 (n))} P(W, \textit{linear})$.
\end{theorem}
The protocol introduced in \cite{Buhrman:2024} uses multiple subroutines. 
\cref{sec:success_subroutines} gives the success probability for each of these subroutines. 
Next, \cref{sec:success_adaptive_W_state,sec:success_standard_W_state} give the success probabilities for preparing the $W$-state using an adaptive approach and using a standard approach. 
Finally, \cref{sec:success_compare_W_state} compares the two approaches and proves the theorem.

\subsection{Success probability for different subroutines}\label{sec:success_subroutines}
The adaptive approach to prepare the $W$-state uses an efficient mapping between the binary number representation and the one-hot number representation.
These mappings use different subroutines to implement the fanout gate and the OR-gate (which evaluates to 1 precisely if the OR of the inputs evaluates to 1). 
The OR-gate, defined by \citeauthor{TakahashiTani:2013}, has exponential size~\cite{TakahashiTani:2013}. 
By first applying the OR-reduction introduced by \citeauthor{HoyerSpalek:2005}~\cite{HoyerSpalek:2005}, we prepare a specific quantum state of logarithmic size.
Evaluating the OR of the logarithmic-sized quantum state yields the same outcome as when evaluating the OR of the original quantum state. 
Combined, the two approaches give a polynomial-sized circuit to compute the OR of a quantum state.

\subsubsection{Fanout and Parity gate}
\begin{figure}[th]
    \centering
    \begin{quantikz}
    \lstick{$\ket{\phi}$}\slice{t=0} & \slice{t=1} & \slice{t=2} & \ctrl{3}\slice{t=3} & \slice{t=4} & \slice{t=5} & & \slice{t=6} & \slice{t=7} & \slice{t=8} & \gate{Z}\slice{t=9} & \\
    \lstick{$\ket{x_1}$} & & & & & & \targ{} & & & & & \\
    \lstick{$\ket{x_2}$} & & & & & & & \targ{} & & & & \\
    \lstick{$\ket{0}$} & \gate{H} & \ctrl{1} & \targ{} & \meter{} & \ctrl[vertical wire=c]{1}\setwiretype{c} \\
    \lstick{$\ket{0}$} & & \targ{} & \targ{} & \meter{} & \ctrl[vertical wire=c]{1}\setwiretype{c} \\
    \lstick{$\ket{0}$} & \gate{H} & \ctrl{1} & \ctrl{-1} & & \targ{} & \ctrl{-4} & & \gate{H} & \meter{} & \ctrl[vertical wire=c]{-5}\setwiretype{c} & \setwiretype{n} \\
    \lstick{$\ket{0}$} & & \targ{} & \targ{} & \meter{} & \ctrl[vertical wire=c]{1}\wire[u][1]{c}\setwiretype{c} & \setwiretype{n} \\
    \lstick{$\ket{0}$} & \gate{H} & & \ctrl{-1} & & \targ{} & & \ctrl{-5} & \gate{H} & \meter{} &  \ctrl[vertical wire=c]{-2}\setwiretype{c} & \setwiretype{n} 
    \end{quantikz}
    \caption{Implementation of a quantum fanout gate with the GHZ state preparation expanded.
    The time steps indicate which gates can be applied in parallel.}
    \label{fig:q_circuit:non_local_cnot_expanded}
\end{figure}
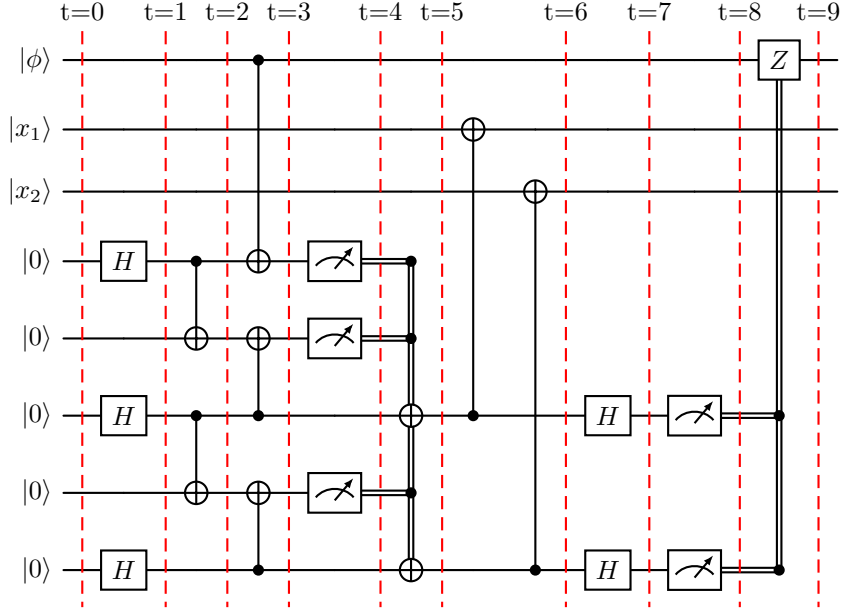

A fanout gate on $n$ qubits has one control qubit and $n-1$ target qubits. 
Our implementation requires $3n-1$ qubits and is inspired by the non-local CNOT-gate by \citeauthor{YimsiriwattanaLomonaco:2004}~\cite{YimsiriwattanaLomonaco:2004}.
The circuit first prepares a GHZ state on $n$ qubits using $2n-1$ qubits and then uses this GHZ state to apply parallel gate teleportation to implement the fanout gate. 
\cref{fig:q_circuit:non_local_cnot_expanded} gives the corresponding circuit for $n=3$, the time steps indicate which gates can be applied in parallel. 
The circuit naturally extends to larger $n$. 

Counting the operations and idling terms gives the success probability. 
We also obtain the success probability for the parity gate, by noting that conjugating the Fanout gate by Hadamard gates yields the Parity gate. 
\begin{align}
    P(Fanout_{n}) & \ge p_{s}^{2n + \ceil{(n-1)/2}} p_{is}^{5n + \floor{(n-1)/2} - 2} p_{d}^{3n - 2} p_{id}^{2n + 1} p_{m}^{2n - 1} p_{im}^{3n - 1} p_{ic}^{3n - 1}. \label{eq:success:fanout} \\
    P(Parity_{n}) & \ge p_{s}^{4n + \ceil{(n-1)/2} - 1} p_{is}^{3n + \floor{(n-1)/2} - 1} p_{d}^{3n - 2} p_{id}^{2n + 1} p_{m}^{2n - 1} p_{im}^{3n - 1} p_{ic}^{3n - 1}. \label{eq:success:parity}
\end{align}
As the circuits for both the fanout gate and the parity gate have the same depth and apply the same type of operations, we see that a qubit idling during the execution of either of the two gates has the same success probability. 
Counting the gates gives a success probability for an idling qubit of
\begin{equation}\label{eq:success:fanout_idle}
    P(iFanout_{n}) = P(iParity_{n}) = p_{is}^{4} p_{id}^{3} p_{im}^{2} p_{ic}^{2}.
\end{equation}

\subsubsection{OR-reduction}
The OR-reduction introduced by \citeauthor{HoyerSpalek:2005} prepares a state on $\bigo(\log n)$ qubits, such that evaluating an OR-gate on this reduced state gives the same output as evaluating the OR-gateon the initial $n$ qubits~\cite[Lemma~5.1]{HoyerSpalek:2005}.

Let $c$ be any positive integer and $\varphi\in [0,2\pi)$, define the state
\begin{equation*}
    \ket{\mu_{\varphi}^{c}} = \frac{1+e^{i\varphi c}}{2}\ket{0} + \frac{1-e^{i\varphi c}}{2}\ket{1}.
\end{equation*}
We obtain this state by computing: $\ket{\mu_{\varphi}^{c}}= HR_Z(\varphi c) H \ket{0}$. 
The OR-reduction uses these $\ket{\mu_{\varphi}^{c}}$ states. 

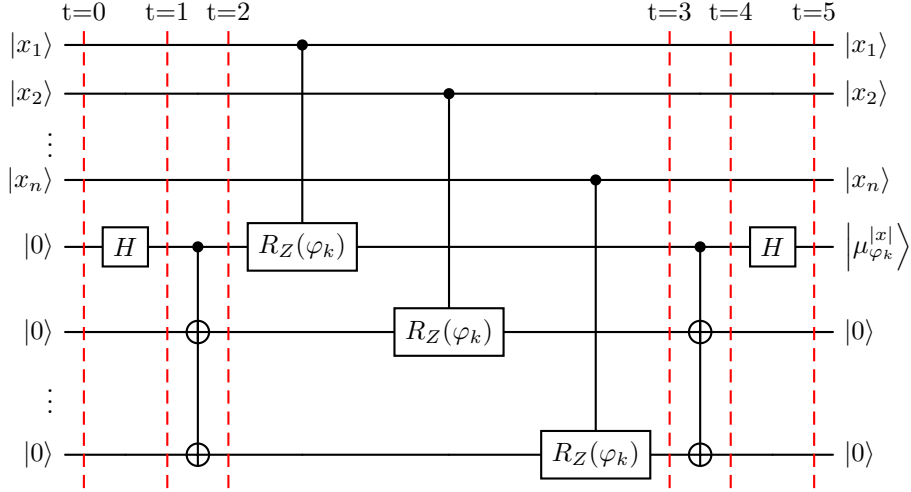
\begin{figure}
    \centering
    \begin{quantikz}
    \lstick{$\ket{x_1}$}\slice{t=0} & \slice{t=1} & \slice{t=2} & \ctrl{4} & & \slice{t=3} & \slice{t=4} & \slice{t=5} & \rstick{$\ket{x_1}$} \\
    \lstick{$\ket{x_2}$} & & & & \ctrl{4} & & & & \rstick{$\ket{x_2}$} \\
    \lstick{$\vdots$} \\
    \lstick{$\ket{x_n}$} & & & & & \ctrl{4} & & & \rstick{$\ket{x_n}$} \\
    \lstick{$\ket{0}$} & \gate{H} & \ctrl{3} & \gate{R_Z(\varphi_k)} & & & \ctrl{3} & \gate{H} & \rstick{$\ket{\mu_{\varphi_k}^{|x|}}$} \\
    \lstick{$\ket{0}$} & & \targ{} & & \gate{R_Z(\varphi_k)} & & \targ{} & & \rstick{$\ket{0}$} \\
    \lstick{$\vdots$} \\
    \lstick{$\ket{0}$} & & \targ{} & & & \gate{R_Z(\varphi_k)} & \targ{} & & \rstick{$\ket{0}$}
    \end{quantikz}
    \caption{An adaptive circuit that prepares the state $\ket{\mu_{\varphi_k}^{|x|}}$. 
    These states are used to implement the OR-reduction. 
    The dotted lines indicate time steps and which gates can be applied in parallel.}
    \label{fig:q_circuit:OR_reduction}
\end{figure}

Given input $x_1,\hdots,x_n$, the OR-reduction prepares $t=\ceil{\log_2 (n+1)}$ states $\ket{\mu_{\varphi_k}^{|x|}}$ for $\varphi_k = \frac{2\pi}{2^k}$ and $k\in [t]$. 
We can prepare the states $\ket{\mu_{\varphi_k}^{|x|}}$ for each $k$ in parallel using fanout gates. 
\cref{fig:q_circuit:OR_reduction} shows the circuit for a single $k$, with the time steps again indicating which gates can be applied in parallel. 

We can parallelize this circuit for every $k$ using fanout gates. 
The $n$ fanout gates of length $t$ for copying the input qubits can be applied in parallel with the $t$ fanout gates of length $n$ for copying the auxiliary qubits. 
As a result, we have no idling terms for the fanout gates. 
Additionally, the first and last Hadamard gate in \cref{fig:q_circuit:OR_reduction} can be included in the construction of the fanout gates. 
This gives a total success probability of: 
\begin{align}
    & P(OR_{n}\text{-}reduction) \nonumber \\
    & \qquad = P(Fanout_{t})^{2n}P(Fanout_{n})^{2t} \big(P(cR_Z)^{n}\big)^{t} \nonumber \\
    & \qquad \ge p_{s}^{11nt + 2(n\ceil{(t-1)/2} + t\ceil{(n-1)/2}) + 2t} p_{is}^{23nt + 2n\floor{(t-1)/2} + 2t\floor{(n-1)/2} - 4(n+t)} \nonumber \\
    & \qquad p_{d}^{14nt - 4(n+t)} p_{id}^{8nt + 2(n+t)} p_{m}^{8nt - 2(n+t)} p_{im}^{12nt - 2(n+t)} p_{ic}^{12nt - 2(n+t)} \label{eq:success:OR_reduction}
\end{align}
We now use this expression to determine the success probability for the OR-gate.

\subsubsection{OR-gate}
\citeauthor{TakahashiTani:2013} presented an exponential-sized circuit for the OR-gate, which they applied to a state of logarithmic size~\cite{TakahashiTani:2013}. 
Combined, this gives a circuit of polynomial size.
From the Fourier inversion formula, we know that for a bit string $x\in\F_2^n$ we can write
\begin{equation*}
    \text{OR}_n(x) = \frac{1}{2^{n-1}}\sum_{a\in\F_2^n\setminus \{0^n\}} \text{PA}_n^a(x),
\end{equation*}
where PA$_n^a(x) = \oplus_{j=0}^{n-1} a_i x_i$ is the parity of $x$, weighted by a nonzero vector~$a$.
This weighted-parity gate is implemented by a standard parity gate on a subset of the inputs.
The exponential-sized circuit for the OR$_n$-gate consists of three steps: 
\begin{enumerate}
    \item Simultaneously, 
    \begin{enumerate}
        \item Copy the input state $2^n-1$ times and compute PA$_n^a(x)$ for every nonzero $n$-bit string $a$;
        \item Prepare a GHZ state on $2^n-1$ qubits;
    \end{enumerate}
    \item Apply $R_Z(\pi/2^{n-1})$-gates to the qubits in the GHZ state and controlled by the qubits that hold the PA$_n^a(x)$ results;
    \item Apply a fanout gate of length $2^n-1$ to the GHZ state together with a Hadamard gate to uncopy the GHZ state and obtain the OR of the input in a single auxiliary qubit. 
\end{enumerate}
We then uncompute the auxiliary registers by rerunning the first two steps.
This introduces an extra factor two in the exponents in the overall success probability. 
Furthermore, in this protocol, we can choose to prepare the GHZ states only when they are needed, thereby omitting idling terms for the GHZ states. 

In total, we will have $n$ parity gates with a single target, $\binom{n}{2}$ parity gates with two targets, and in general $\binom{n}{k}$ parity gates with $k$ targets, for $k$ the Hamming weight of $a$. 
By replacing weighted-parity gates by the success probability of the parity gate on all inputs, we obtain a simpler lower bound on the success probability.

Let $n$ be the length of the input and $t=\ceil{\log_2(n+1)}$. 
The success probability for the OR-gate is then given by
\begin{align}
    & P(OR_{n}) \nonumber \\
    & = P(OR_{n}\text{-}reduction)^{2} P(Fanout_{2^{t-1}})^{2t} \left(\Pi_{k=1}^{t} \binom{t}{k} P(Parity_{k})\right)^{2} \nonumber \\
    & \qquad P(GHZ_{2^t-1}, \textit{adaptive})^{2} P(cR_Z)^{2^{t}-1} p_{s}^{1} \nonumber \\
    & \ge P(OR_{n}\text{-}reduction)^{2} P(Fanout_{2^{t-1}})^{2t} P(Parity_{t})^{2(2^{t}-1)} \nonumber \\
    & \qquad P(GHZ_{2^t-1}, \textit{adaptive})^{2} P(cR_Z)^{2^{t}-1} p_{s}^{1}
     \nonumber \\
    & \ge p_{s}^{22nt + 2(2n - 2^{t} - 1)\ceil{(t-1)/2} + 4t\ceil{(n-1)/2} + 2t\ceil{(2^{t-1}-1)/2} + 2\ceil{(2^t-1)/2} + 10t\cdot 2^{t} + 3\cdot 2^{t} - 4t - 2} \nonumber \\
    & \qquad p_{is}^{2(2n + 2^{t} - 1)\floor{(t-1)/2} + 4t\floor{(n-1)/2} + 2t\floor{(2^{t-1}-1)/2} + 2\floor{(2^t-1)/2}} \nonumber \\
    & \qquad p_{is}^{46nt - 8n - 18t + 11t\cdot 2^{t} + 3\cdot 2^{t} - 5} p_{d}^{28nt - 8n - 18t + 9t\cdot 2^{t} + 2\cdot 2^{t} - 6} p_{id}^{16nt + 4n + 2t + 6t\cdot 2^{t} + 2\cdot 2^{t} + 2} \nonumber \\
    & \qquad p_{m}^{16nt - 4n - 10t + 6t\cdot 2^{t} - 2} p_{im}^{24nt - 4n - 12t + 9t\cdot 2^{t}} p_{ic}^{24nt - 4n - 12t + 9t\cdot 2^{t}}. \label{eq:success_OR_gate_exact}
\end{align}
If $n=2^k$, the expression simplifies, as in that case $t=\ceil{\log_2(n+1)} = k+1$, giving
\begin{align}
    P(OR_{n}) & \ge p_{s}^{42nk + 50n - 4k + 2(2k - 2n + 1)\ceil{k/2} + 6(k+1)\ceil{(n-1)/2} - 6} \nonumber \\
    & \qquad p_{is}^{68nk + 68n - 18k + 2(4n - 1)\floor{k/2} + 6(k+1)\floor{(n-1)/2} - 25} p_{d}^{46nk + 42n - 18k - 24} \nonumber \\
    & \qquad p_{id}^{28nk + 36n + 2k + 4} p_{m}^{28nk + 24n - 10k - 12} p_{im}^{42nk + 38n - 12k - 12} p_{ic}^{42nk + 38n - 12k - 12}. \label{eq:success_OR_gate_simplified}
\end{align}

\subsection{Adaptive approach}\label{sec:success_adaptive_W_state}
This section gives the success probability for preparing the $W$-state on $n=2^k$ qubits, for some integer $k$, using the approach introduced in~\cite{Buhrman:2024}. 
Their approach uses a compress-uncompress technique to map between different number representations:
\begin{align}
    \text{\textbf{Uncompress}: }& \ket{i}_{\log n}\ket{0}_{n} \mapsto \ket{i}_{\log n}\ket{e_i}_{n}, \label{eq:unitary:uncompress} \\
    \text{\textbf{Compress}: }& \ket{i}_{\log n}\ket{e_i}_{n} \mapsto \ket{0}_{\log n}\ket{e_i}_{n}. \label{eq:unitary:compress}
\end{align}
The \textbf{Uncompress} method uses Equal$_i$ gates, that evaluate to $1$ precisely if the input state corresponds to the $i$-th computational basis state, to set the qubits in the target register. 
Next, the \textbf{Compress} method cleans the logarithmic-sized register using controlled-$Z$ gates. 
Fanout-gates make sure that we can implement all these operations simultaneously. 
We refer the reader to~\cite{Buhrman:2024} for details on the exact implementation. 

Following the original circuit, we derive success probabilities for the two methods \textbf{Uncompress} and \textbf{Compress} methods.
\begin{align}
    P(\textbf{Uncompress}_n) & = p_{s}^{k} p_{is}^{nk + n - k} P(Fanout_{n})^{2k} P(iFanout)^{2n} \big(\Pi_{i=0}^{n-1} P(Equal_{i})\big). \label{eq:success_uncompress} \\
    P(\textbf{Compress}_n) & = p_{s}^{2k} p_{is}^{2(nk + n - k)} P(Fanout_{n})^{2k} P(iFanout)^{2n} \big(\Pi_{i=0}^{n-1} P(cZ, target_{i})\big). \label{eq:success_compress}
\end{align}

The success probability of the Equal$_i$-gate is lower bounded by the success probability of the OR$_k$-gate with all $k$ input qubits conjugated with $X$-gates. 
We can incorporate these $X$-gates in the circuit for the OR$_k$-gate. 
The controlled-$Z$-gates with target $i$ correspond to a fanout gate with targets on the qubits corresponding to the ones in the binary representation of $i$, and with the targets conjugated by Hadamard gates. 
Hence, we can lower bound the success probability of these controlled-$Z$-gates with target $i$ by the success probability of a fanout gate of length $k+1$ with the target qubits conjugated by Hadamard gates. 
Summarizing, we have that for every $i\in\F_{2}^{k}$
\begin{align}
    P(Equal_{i}) & \ge P(OR_{k})p_{s}^{2k}p_{is}^{2}, \label{eq:success_equal_i} \\
    P(cZ, target_{i}) & \ge P(Fanout_{k+1})p_{s}^{2k}p_{is}^{2}. \label{eq:success_cZ_target}
\end{align}

We can now obtain a lower bound on the success probability of preparing the $W$-state by multiplying the expressions of \cref{eq:success_uncompress,eq:success_compress}, applying the lower bounds described in \cref{eq:success_equal_i,eq:success_cZ_target} and using the expression for the OR-gate given in \cref{eq:success_OR_gate_exact}.
We use $t=\ceil{\log_2(k+1)}$ and obtain
\begin{align}
    & P(W,\textit{adaptive}) = P(\textbf{Uncompress}_n) P(\textbf{Compress}_n) \nonumber \\
    & \enspace \ge p_{s}^{4nk + 3k} p_{is}^{3nk + 7n - 3k} P_{Fanout_{n}}^{4k} P_{Fanout_{k+1}}^{n} P_{iFanout}^{4n} P_{OR_{k}}^{n} \nonumber \\
    & \enspace \ge p_{s}^{22nkt + 14nk + 2n\ceil{(2^t-1)/2} + n(3\cdot 2^{t} + \ceil{k/2}) + 2nt(5\cdot 2^{t} - 2) + 3k + 4k\ceil{(n-1)/2} + 2n(2k - 2^{t} - 1)\ceil{(t-1)/2}} \nonumber \\
    & \qquad p_{s}^{4nt\ceil{(k-1)/2} + 2nt\ceil{(2^{t-1}-1)/2} + 2n\ceil{(2^t-1)/2}} p_{is}^{46nkt + 20nk + 3n\cdot 2^{t} - 18nt + 21n - 11k + n\floor{k/2}} \nonumber \\
    & \qquad p_{is}^{2n(2k + 2^{t} - 1)\floor{(t-1)/2} + 4nt\floor{(k-1)/2} + 2nt\floor{(2^{t-1}-1)/2} + 2n\floor{(2^t-1)/2} + 11nt\cdot 2^{t} + 4k\floor{(n-1)/2}} \nonumber \\
    & \qquad p_{d}^{28nkt + 7nk + 9nt(2^{t} - 2) + 2n\cdot 2^{t} - 5n - 8k} p_{id}^{16nkt + 14nk + 2nt(3\cdot 2^{t} + 1) + 2n\cdot 2^{t} + 17n + 4k} \nonumber \\
    & \qquad p_{m}^{16nkt + 6nk + 2nt (3\cdot 2^{t} - 5) - n - 4k} (p_{im}p_{ic})^{24nkt + 11nk + 3nt(3\cdot 2^{t} - 4) + 10n - 4k}. \label{eq:success_W_adaptive_exact} 
\end{align}
Note that this expression is quite involved with dependencies on both $n$, $k=\log_2(n)$ and $t=\ceil{\log_2(k+1)}$. 
We furthermore use ceil- and floor-functions. 
We can approximate \cref{eq:success_W_adaptive_exact} using $\ceil{x}\approx x\approx \floor{x}$ for $x\in \R$ and $2^t \approx k$. 
We then obtain
\begin{align}
    P(W,\textit{adaptive}) & \gtrsim p_{s}^{71nkt/2 + 37nk/2 + 3nk - 8nt - n + k} p_{is}^{125nkt/2 + 47nk/2 - 22nt + 21n - 13k} \nonumber \\
    & \qquad p_{d}^{37nkt + 9nk - 18nt - 5n - 8k} p_{id}^{22nkt + 16nk + 2nt + 17n + 4k} \nonumber \\
    & \qquad p_{m}^{22nkt + 6nk - 10nt - n - 4k} (p_{im}p_{ic})^{33nkt + 11nk - 12nt + 10n - 4k}. \label{eq:success_W_adaptive_approximate} 
\end{align}

\subsection{Non-adaptive method}\label{sec:success_standard_W_state}
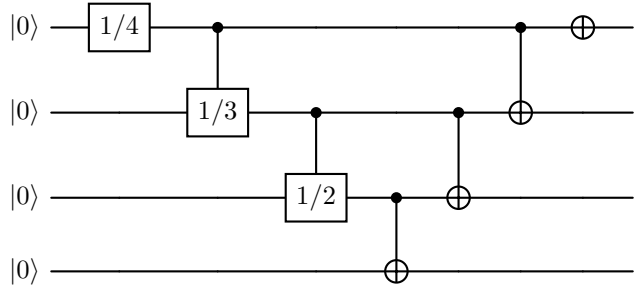
\begin{figure}
    \centering
    \begin{quantikz}
    \lstick{\ket{0}} & \gate{1/4} & \ctrl{1} & & & & \ctrl{1} & \targ{} & \\
    \lstick{\ket{0}} & & \gate{1/3} & \ctrl{1} & & \ctrl{1} & \targ{} & & \\
    \lstick{\ket{0}} & & & \gate{1/2} & \ctrl{1} & \targ{} & & & \\
    \lstick{\ket{0}} & & & & \targ{} & & & &
    \end{quantikz}
    \caption{Exact circuit for preparing the $W$-state for $n=4$.
    Every gate parametrized by $1/n$ denotes a controlled-$R_Y$-gate with argument $\theta = -2\arccos{\sqrt{1/n}}$.}
    \label{fig:q_circuit:W_prep:exact}
\end{figure}

A non-adaptive method of preparing a $W$-state uses successive controlled-$R_Y$-gates, where the angle of the gates depends on the qubit index. 
\cref{fig:q_circuit:W_prep:exact} shows the circuit for $n=4$. 
Each square $1/n$ denotes an $R_Y(\theta)$-gate with argument $\theta = -2\arccos{\sqrt{1/n}}$.
The controlled-$R_Y$-gate reduce to a CNOT-gate for $n=1$.

The quantum circuit shown in \cref{fig:q_circuit:W_prep:exact} naturally extends to arbitrary $n$ and prepares the $W$-state on $n$ qubits. 
Just before the first CNOT-gate, the quantum circuit corresponds to the state 
\begin{equation*}
    \frac{1}{\sqrt{n}} \sum_{i=0}^{n-1} \ket{1}^{i}\ket{0}^{n-i}.
\end{equation*}
Each CNOT-gate will then correctly set one additional qubit, until we have the desired $W$-state. 
The idea behind this circuit is to iteratively ``pass on'' part of the amplitude to the remaining unset qubits and thereby correctly set all qubits. 

\begin{figure}
    \centering
    \begin{quantikz}
    \lstick{\ket{0}} & \gate{1/4} & \ctrl{1} & & \ctrl{1} & & & & & & & & \ctrl{1} & \targ{} & \\
    \lstick{\ket{0}} & \gate{C} & \targ{} & \gate{B} & \targ{} & \gate{A} & \ctrl{1} & & \ctrl{1} & & & \ctrl{1} & \targ{} & &  \\
    \lstick{\ket{0}} & & & & & \gate{C'} & \targ{} & \gate{B'} & \targ{} & \gate{A'} & \ctrl{1} & \targ{} & & & \\
    \lstick{\ket{0}} & & & & & & & & & & \targ{} & & & & 
    \end{quantikz}
    \caption{Exact decomposed circuit for preparing the $W$-state for $n=4$, where every controlled $R_Y$-gate is replaced by single-qubit gates and CNOT-gates.}
    \label{fig:q_circuit:W_prep:exact_decomposed}
\end{figure}
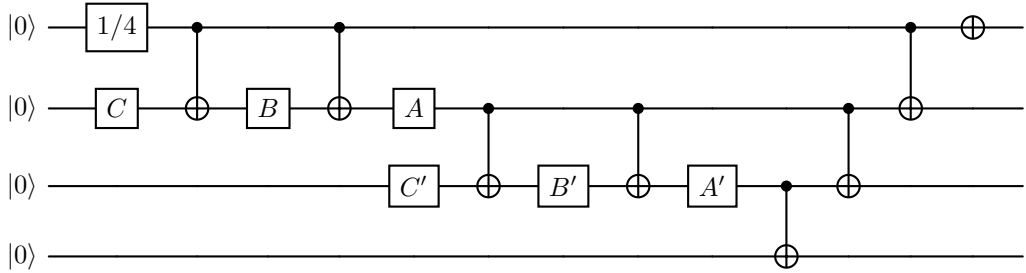

The circuit shown in \cref{fig:q_circuit:W_prep:exact} uses controlled-$R_Y$-gates, which most quantum hardware devices do not directly support. 
We decompose the controlled-$R_Y$-gates in a circuit with three single-qubit gates and CNOT-gates in between. 
In our case, the exact single-qubit gates used in the decomposition are irrelevant, as we assume the same success probability for every single-qubit gate. 
\cref{fig:q_circuit:W_prep:exact_decomposed} shows the resulting decomposed quantum circuit to prepare the $W$-state for $n=4$. 

This figure shows that some gates can be applied in parallel. 
Counting the depth gives $n-2$ groups of four layers each, with two single-qubit gates, a CNOT-gate, one single-qubit gate and again a CNOT-gate. 
Next, there is one single-qubit gate, $n-1$ layers of CNOT-gates and a final single-qubit gate, for a total depth of $5n-7$ for $n\ge 2$. 
The total success probability is given by
\begin{equation} \label{eq:success_W_direct}
    P(W,\textit{non-adaptive}) = p_{s}^{3n - 4} p_{is}^{n(2n - 5) + 4} p_{d}^{3n - 5} p_{id}^{n(3n - 11) + 10}.
\end{equation}

\subsection{Comparison success probability}\label{sec:success_compare_W_state}
To determine when the adaptive approach outperforms the standard approach, we have to determine when 
\begin{equation}
    P(W,\textit{adaptive}) \ge P(W, \textit{non-adaptive}).
\end{equation}
Comparing the success probabilities given in \cref{eq:success_W_adaptive_exact,eq:success_W_direct} shows that this inequality holds approximately if the following inequality holds
\begin{align*}
    & p_{s}^{71nkt/2 + 37nk/2 + 3nk - 8nt - 4n + k + 4} p_{d}^{37nkt + 9nk - 18nt - 8n - 8k + 5} \\
    & p_{m}^{22nkt + 6nk - 10nt - n - 4k} (p_{im}p_{ic})^{33nkt + 11nk - 12nt + 10n - 4k} \\
    & \enspace \ge p_{is}^{2n^2 - 125nkt/2 - 47nk/2 + 22nt - 26n + 13k + 4} p_{id}^{3n^2 - 22nkt - 16nk - 2nt - 28n - 4k + 10}.
\end{align*}
Applying the assumptions on the success probabilities discussed in \cref{sec:error_model}, shows that this inequality reduces to 
\begin{equation*}
    p_{d}^{59nkt + 15nk - 28nt - 9n - 12k + 5} \gtrsim p_{id}^{3n^2 - 88nkt - 38nk + 22nt - 48n + 4k + 10}.
\end{equation*}
For a first estimate on when the adaptive approach outperforms the standard direct approach, we only keep the most significant terms.
Using that $k=\log_2 n$ and $t\approx \log_2 \log_2 n$ shows that the adaptive approach performs best if
\begin{equation*}
    p_{d} \gtrsim p_{id}^{3n/(59\log_2 n \log_2\log_2 n)}.
\end{equation*}
Let $\eps>0$, now if $p_d = (1+\eps)p_{id}^{3n/(59\log_2 n \log_2\log_2 n)}$, then we see that 
\begin{equation*}
P(W,\textit{adaptive}) \gtrsim (1+\eps)^{59n\log_2 n\log_2\log_2 n} P(W, \textit{non-adaptive}),    
\end{equation*}
completing the proof of \cref{thm:error:W_state}.

Looking at the time aspect of the two approaches, we see that the time required to run the non-adaptive quantum protocol scales linearly in $nt_d$, where $n$ is the size of the $W$-state and $t_d$ the duration of a two-qubit gate. 
The adaptive quantum protocol uses a constant number of consecutive two-qubit gates, measurements and intermediate classical computations. 
As long as these times remain constant, the adaptive protocol will be faster. 
Note that the classical computations take $\bigo(\log n)$ time.
However, these operations are typically significantly faster than quantum gates, which mitigates the effect of scaling computation times. 
Additionally, the scaling remain logarithmic versus linear in the adaptive algorithm, giving an exponential advantage over non-adaptive algorithms. 

\section{Discussion}\label{sec:discussion}
This work analyzed the potential of adaptive quantum computers to prepare quantum states and outperform non-adaptive quantum alternatives. 
We first derived expressions that compute the success probability of various quantum state preparation protocols (both adapative and non-adaptive) and then compared them to learn when protocol outperforms another. 
Next, we also implemented protocols to prepare the GHZ state to see how well the theory aligns with practice.

We found that in practice, the protocols perform worse than we might expect from the theoretical derivations. 
Additionally, the non-adaptive approach seems to work better, even when the derived formulas predict otherwise. 
Many reasons exist that (partially) explain this observation. 
First, the error model used was worst case, where no error was allowed and additionally no two errors were expected to cancel each other. 
In practice, we see that some errors are worse than others, and errors might even cancel. 
Furthermore, our implementation only allowed us to measure bit errors in the final state. 
This error model furthermore assumed all single-qubit gates had the same success probability.
In practice however, gates have to be decomposed in terms of the available gate set, often resulting in overhead. 

Second, the implementation differed from the original protocol. 
The main reason was the limited use of intermediate classical computations. 
The hardware backend only allowed for classical control of future quantum gates. 
The protocol however wished to perform a prefix parity computation on the measurement outcomes. 
We have implemented this by instead determining the required single-qubit gates for every measurement outcome individually, resulting in significant overhead. 
We expect that over time, the possibilities for intermediate classical computations will improve, opening the way to improved adaptive quantum algorithms. 

Third, the implementations might have room for improvement.
With full control over the quantum computer, hardware-specific optimizations can be used to improve the overall implementation of the algorithm. 
We instead used the standard programming environment to formulate our algorithm and did not tailor our implementation to the specific hardware backend. 
Additionally, the compiler was used as black-box to ensure that the algorithm could be run on the quantum hardware. 
Higher optimization levels could simplify and improve the implementation further. 

Fourth, we assumed no limitation on the number of parallel gates. 
The adaptive approach uses a dense quantum circuit, where qubits are idle only for short periods of time. 
However, current devices often have a limit on the number of gates that can be applied in parallel, for instance to minimize the effect of crosstalk~\cite{Zhao:2022,Zhou:2023}. 
Physically separating qubits can help in reducing crosstalk, but also introduces new challenges, such as letting two distant qubits interact easily. 
The additional idling times imply that the adaptive quantum circuits are less dense than expected, giving more room for qubits to decohere before the end of the circuit. 

In our implementations, we saw that the gate count and circuit size are good predictors for the expected improvement of one method over another. 
We saw that the non-adaptive implementation resulted in a wider spread of the probability distribution. 
The differences with the outcomes of the adaptive approach became more prominent for larger $n$. 
We expect that the idling times in the adaptive circuit were longer in practice than in the theoretical derivations, due to the reasons outlined before, thus resulting in more qubits decohering than expected. 

For future research, we envision two main directions. 
The first direction extends this work by enriching the error model used. 
A different error model might better describe the actual behavior of the quantum system, however would also complicate the analysis. 
These different error models might also drop initial assumptions on the independence of errors or include error mitigation techniques. 

The second direction allows for a broader class of algorithms. 
For instance, algorithms to prepare other quantum states, or algorithms that have access to a limited gate set. 
This limited gate set requires the algorithms to prepare the quantum state up to some error with respect to some norm. 
Similarly, we can explore probabilistic algorithms that only output the correct state with certain probability.
A simple probabilistic algorithm that approximately prepares the $W$-state is to apply $n$ single-qubit $R_Y$-gates with parameter $\theta=\arccos\big(\sqrt{\tfrac{n-1}{n}}\big)$ and then measuring the parity of these gates.
Upon measuring an odd parity, the superposition collapses to a superposition over all bit strings of odd Hamming weight, with those having Hamming weight~$1$ having the highest amplitudes. 
Choosing a smaller $\theta$ reduce the probability of finding an odd parity. 
However, once an odd parity is measured, the resulting state better approximates the $W$-state. 

\printbibliography

\end{document}